\documentclass[review]{elsarticle}

\usepackage{bm}
\usepackage{mathrsfs}
\usepackage{subfigure}
\usepackage{booktabs}
\usepackage{wrapfig}
\usepackage{picins}
\usepackage{amssymb}
\usepackage{amsmath}
\usepackage{graphicx}
\usepackage{epstopdf}
\usepackage{color}
\usepackage{multirow}
\usepackage{geometry}
\journal{}

\begin{document}
    \pagestyle{empty}
    \begin{frontmatter}

\title{A survey on active noise control techniques--Part I: Linear systems}

\tnotetext[mytitlenote] {The work is supported by the National Science Foundation of P.R. China under Grant 61901285, 61901400, and 61701327, Sichuan Science and Technology Fund under Grant 20YYJC3709, China Postdoctoral Science Foundation under Grant 2020T130453, and Sichuan University Postdoctoral Interdisciplinary Fund.\\E-mail\;addresses: \texttt{lulu19900303@126.com(L.\;Lu), kl\_yin@hotmail.com(K.-L.\;Yin), delamare@cetuc.puc-rio.br(R.C.\;de\;Lamare), zongsheng56@126.com(Z.\;Zheng),  yuyi\_xyuan@163.com(Y.\;Yu), arielyang@scu.edu.cn(X. Yang), chenbd@mail.xjtu.edu.cn(B. Chen)}. \\Corresponding author: Xiaomin Yang.}

\author{Lu Lu$^{a}$, Kai-Li Yin$^{b}$, Rodrigo C. de Lamare$^{c}$, Zongsheng Zheng$^{d}$, Yi Yu$^{e}$, Xiaomin Yang$^{a*}$, Badong Chen$^{f}$}
\address{a) School of Electronics and Information Engineering, Sichuan University, Chengdu, Sichuan 610065, China.}
\address{b) School of Computer Science, Sichuan University, Chengdu, Sichuan 610065, China.}
\address{c) CETUC, PUC-Rio, Rio de Janeiro 22451-900, Brazil.}
\address{d) School of Electrical Engineering, Sichuan University, Chengdu, 610065, China.}
\address{e) School of Information Engineering, Robot Technology Used for Special Environment Key Laboratory of Sichuan Province, Southwest University of Science
and Technology, Mianyang 621010, China.}
\address{f) School of Electronic and Information Engineering, Xi'an Jiaotong University, Xi'an 710049, China.}

\begin{abstract}
Active noise control (ANC) is an effective way for reducing the noise level in electroacoustic or electromechanical systems. Since its first introduction in 1936, this approach has been greatly developed. This paper focuses on discussing the development of ANC techniques over the past decade. Linear ANC algorithms, including the celebrated filtered-x least-mean-square (FxLMS)-based algorithms and distributed ANC algorithms, are investigated and evaluated. Nonlinear ANC (NLANC) techniques, such as functional link artificial neural network (FLANN)-based algorithms, are pursued in Part II. Furthermore, some novel methods and applications of ANC emerging in the past decade are summarized. Finally, future research challenges regarding the ANC technique are discussed.
\end{abstract}


\end{frontmatter}


\section{Introduction}
\label{sec:Introduction}

Traditional noise control methods use passive noise control (PNC)
techniques such as sealing and shielding to attenuate noise.
However, such methods suffer from the bulky volume, expensive cost,
and degradation performance for reducing low-frequency noise, which
hinders its practical use \cite{hansen2012active}. To address this
problem, active noise control (ANC) was developed, based on the
principle that noise can be canceled by another noise with the same
amplitude but an opposite phase
\cite{nelson1991active,kuo1996active}. The development of ANC
history can be found in \cite{elliott1993active,george2013advances}
and references therein.

\begin{figure}[!htb]
    \centering
    \includegraphics[scale=0.5]{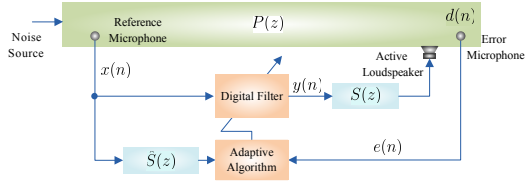}
    \caption{\label{01} Diagram of the feedforward ANC model, where $n$ is the iteration, and $z$ is for z-transform.}
    \label{Fig01}
\end{figure}

In ANC, the most popular adaptive algorithm is the filtered-x least-mean-square (FxLMS) algorithm. The FxLMS algorithm exhibits a simple structure, thus it has been extensively studied and extended \cite{lorente2015the,shi2019two-gradient,tufail2019two}. Fig. \ref{Fig01} plots the diagram of a feedforward ANC model, where $P(z)$ denotes the primary path, $S(z)$ denotes the secondary path, which can be used to model the acoustic path between the loudspeaker and the microphone, or the electro-acoustic path that also includes the effects of the amplifier and the driver circuit, $\hat S(z)$ represents the estimate of the secondary path model, $x(n)$ stands for the reference signal, $d(n)$ denotes the undesired signal, $y(n)$ denotes the output of the controller, and $e(n)$ denotes the residual  noise. Morgan's experiment demonstrated that, for the narrowband ANC (NANC) system, the convergence property of the adaptive algorithm largely relies on the phase response of the filter $S(z)$ \cite{morgan2013history}. As the phase increases, it will oscillate and eventually cause instability in the entire ANC system. An effective solution is the introduction of the estimated secondary path, i.e., $\hat S(z)$. The above method is usually referred to as \textit{filtered} approach. However, the secondary path increases its eigenvalue spread, and it slows down the convergence speed of the FxLMS algorithm \cite{krstajic2013increasing,zecevic2015new}. More importantly, $\hat S(z)$ is not equal to $S(z)$ in practical applications. Numerous online secondary path estimation methods were developed, which exhibit improved modeling performance than the conventional methods \cite{davari2009designing,wang2009convergence,ardekani2012effects,chang2014feedforward,padhi2018performance,pradhan20205}. In particular, a secondary path modeling for NANC systems was presented in \cite{chang2018secondary}, which analyzed both online and offline estimation methods and demonstrated improved modeling accuracy.

\begin{figure}[!htb]
    \centering
    \includegraphics[scale=0.5]{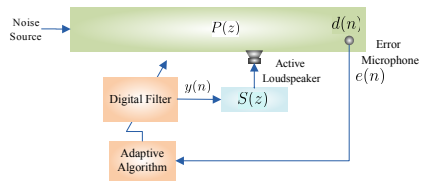}
    \caption{\label{02} Diagram of the feedback ANC model.}
    \label{Fig02}
\end{figure}

Another ANC system model employs the feedback strategy. In contrast with the feedforward ANC system, the feedback ANC system does not need the \textit{a priori} information picked up by the reference microphone, and the attenuated noise level only depends on the active loudspeaker, the adaptive controller, and the error microphone \cite{wu2014a}. Moreover, it is not affected by multiple noise sources \cite{luo2017a}. The block diagram of the feedback ANC system is shown in Fig. \ref{Fig02}. The feedback structure delivers a significantly lower implementation cost, but its drawbacks are also obvious. The main drawback is the stability problem, similar to the  infinite impulse response (IIR) filter. The second weakness is the `waterbed effect' which implies that it is theoretically impossible to suppress noise simultaneously at all frequencies. If the noise at some frequencies is suppressed in a feedback ANC system, the noise will be increased at some other frequencies \cite{wu2018a}. Furthermore, the noise attenuation bandwidth is typically limited. Thus, few feedback ANC systems involve controlling broadband noise, such as chaotic noise and random noise \cite{wu2014a,luo2017a}. A vast number of efforts have been developed to solve these limitations, e.g., see \cite{wu2014a,wu2018a,luo2017a}.

\begin{figure}[h]
    \centering
    \includegraphics[scale=0.5]{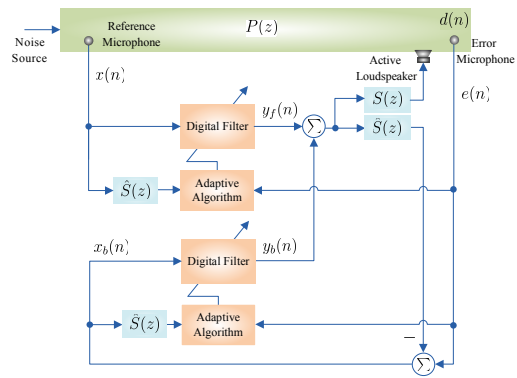}
    \caption{\label{03} Diagram of hybrid ANC model.}
    \label{Fig03}
\end{figure}

The hybrid ANC model combined the feedforward and feedback
structures, whose secondary signal is generated by the sum of the
output of the feedforward and feedback structures. Fig. \ref{Fig03}
shows the diagram of a hybrid ANC model, where $x_b(n)$ and $y_b(n)$
are the reference signal and output of controller in the feedback
structure, respectively, and $y_f(n)$ is the output of the
controller in the feedforward structure. Such model has high design
flexibility, and it can control noise and uncorrelated narrowband
interference generated by other apparatuses
\cite{akhtar2011improving,wu2015decoupling}. A guideline for
selecting an ANC type was proposed in \cite{milani2010maximum},
which analyzes the maximum achievable noise attenuation level for
feed-forward, feedback, and hybrid ANC structures. Over the past
decade, many efforts were conducted by using the hybrid ANC model
\cite{padhi2019new,wang2012psychoacoustic,padhi2017design,padhi2018performance}.
In what follows, we will cover these works in different categories.

Following a different direction, some sparsity-aware ANC algorithms
were proposed to exploit the sparsity of the physical system
\cite{albu2014sparsity}. The convex combination scheme was also
developed for ANC to avert the conflicting requirement between fast
convergence and small residue \cite{ferrer2013convex}. It turns out
that this strategy can be applied not only to linear multi-channel
ANC situations \cite{ferrer2013convex}, but also to nonlinear ANC
(NLANC) systems \cite{george2014convex}. The ANC can cancel the
noise at the microphone location. As a result of the reduction of
the noise level at this point, a spatial zone of quiet (ZoQ) is
created around it. However, its zones of interest are no more than a
finite number of discrete points, leading to a restriction to
generate 3-D quiet zones. In 2013, an ANC algorithm in 3-D space was
developed, which can apply to quiet zones with comparable complexity
and fills the gap in this technology \cite{ardekani2014active}.
Benefiting from wireless acoustic sensor networks (WASNs), an ANC
system over a network of distributed acoustic nodes was proposed,
which is based on \textit{incremental collaborative strategy} with a
sample-by-sample data acquisition in the time-domain
\cite{ferrer2015active}. Following this work, several distributed
algorithms were presented in recent years.

Survey articles on ANC techniques have been published by many
researchers
\cite{george2013advances,kuo2010active,KajikawaRecent,jiang2018review,kuo1999active}.
However, these surveys only focus on one of the problems in ANC, and
they do not cover the literature since 2013. To complete the review
of ANC techniques and include the latest developments, a
comprehensive review from 2009 to 2020 on linear ANC\footnote{To
enhance readability, we also cover the paper published in 2020.},
NLANC, and recent methods and applications are compiled in this
article and the accompanying Part II. In particular, we summarize
and compare the novel modeling methods and algorithms in the last
ten years.

In this Part I, we focus on the development of the past decade of
linear ANC techniques, while in Part II we summarize the development
of NLANC technique and the recent applications of ANC technique. The
paper is organized as follows. In Section \ref{sec:2}, we review the
finite impulse response (FIR) and IIR filter-based ANC algorithms.
In Section \ref{Pra}, some practical considerations in linear ANC
systems are reviewed. In Section \ref{sec:4}, we emphasize the novel
linear ANC methods that emerged in the past decade. Finally, we
summarize the conclusions in Section \ref{sec:5}.

\section{FIR and IIR filter-based ANC algorithms}
\label{sec:2}

Linear ANC systems based on FIR and IIR filters have been
extensively studied in the past decade. This section reviews the
filtered-x-, filtered-e-, and filtered-u-based algorithms, which are
related to standard adaptive algorithms
\cite{jidf,spa,intadap,mbdf,jio,jiols,jiomimo,sjidf,ccmmwf,tds,mfdf,l1stap,mberdf,jio_lcmv,locsme,smtvb,ccmrls,dce,itic,jiostap,aifir,ccmmimo,vsscmv,bfidd,mbsic,wlmwf,bbprec,okspme,rdrcb,smce,armo,wljio,saap,vfap,saalt,mcg,sintprec,stmfdf,1bitidd,jpais,did,rrmber,memd,jiodf,baplnc,als,vssccm,doaalrd,jidfecho,dcg,rccm,ccmavf,mberrr,damdc,smjio,saabf,arh,lsomp,jrpaalt,smccm,vssccm2,vffccm,sor,aaidd,lrcc,kaesprit,lcdcd,smbeam,ccmjio,wlccm,dlmme,listmtc,smcg},
for different types of undesired signals.

\begin{table}[]
        \scriptsize
        \centering \doublerulesep=0.05pt
        \caption{Time line of filtered-x ANC algorithms.}
        \begin{tabular}{l|l|l|l}
            \hline
             \textbf{Years} & \textbf{Authors} & \textbf{Contributions} & \textbf{References} \\ \hline
            1987 &\begin{tabular}[c]{@{}l@{}} Elliott, Stothers,\\ and Nelson \end{tabular} &\begin{tabular}[c]{@{}l@{}}Generalize FxLMS for active control of sound\\ and vibration algorithm\end{tabular} &\begin{tabular}[c]{@{}l@{}} \cite{elliott1987multiple}\end{tabular} \\ \hline
            1992 &\begin{tabular}[c]{@{}l@{}} Shen and\\ Spanias \end{tabular} &\begin{tabular}[c]{@{}l@{}}Frequency-domain FxLMS algorithm for ANC\end{tabular} &\begin{tabular}[c]{@{}l@{}} \cite{shen1992time}\end{tabular} \\ \hline
            1993 &\begin{tabular}[c]{@{}l@{}} Kuo and\\ Luan \end{tabular} &\begin{tabular}[c]{@{}l@{}}Cross-coupled filtered-x LMS algorithm with lattice\\ structure to decorrelate the reference input\end{tabular} &\begin{tabular}[c]{@{}l@{}} \cite{kuo1993cross}\end{tabular} \\ \hline
            1993       &\begin{tabular}[c]{@{}l@{}} Thi and\\ Morgan\end{tabular}         &\begin{tabular}[c]{@{}l@{}}Delayless subband algorithm for ANC\\ \end{tabular}             &\begin{tabular}[c]{@{}l@{}} \cite{thi1993delayless}\end{tabular}            \\ \hline
            2000       &\begin{tabular}[c]{@{}l@{}} Bouchard and\\ Quednau \end{tabular}         &\begin{tabular}[c]{@{}l@{}}Multi-channel filtered-x recursive least square (FxRLS)\\ and multi-channel filtered-x fast-transversal-filter (FTF)\\ algorithms \end{tabular}             &\begin{tabular}[c]{@{}l@{}} \cite{bouchard2000}\end{tabular}            \\ \hline
            2001       &\begin{tabular}[c]{@{}l@{}} Park et al. \end{tabular}         &\begin{tabular}[c]{@{}l@{}}Low-cost delayless subband algorithm\\ \end{tabular}             &\begin{tabular}[c]{@{}l@{}} \cite{park2001delayless}\end{tabular}            \\ \hline
            2003       &\begin{tabular}[c]{@{}l@{}} Lan, Zhang,\\and Ser \end{tabular}         &\begin{tabular}[c]{@{}l@{}} Weight-constrained filtered-x LMS
algorithm\\ for broadband noises\end{tabular}             &\begin{tabular}[c]{@{}l@{}} \cite{wang2003new}\end{tabular}            \\ \hline
            2005       &\begin{tabular}[c]{@{}l@{}}Tobias and\\ Seara\end{tabular}         &\begin{tabular}[c]{@{}l@{}}Analysis of the leaky FxLMS algorithm for Gaussian\\ data and secondary path modeling error \end{tabular}               &\begin{tabular}[c]{@{}l@{}}  \cite{tobias2005leaky} \end{tabular}           \\ \hline
            2005 &\begin{tabular}[c]{@{}l@{}}Carini and\\Sicuranza\end{tabular}   &\begin{tabular}[c]{@{}l@{}}Multi-channel filtered-x affine projection (FxAP)\\ algorithm and its performance analysis\end{tabular}  &\begin{tabular}[c]{@{}l@{}}\cite{carini2005steady}\end{tabular} \\ \hline
            2006       &\begin{tabular}[c]{@{}l@{}} Sun, Kuo,\\and Meng \end{tabular}
            &\begin{tabular}[c]{@{}l@{}}FxLMS algorithm with clipped reference signals for\\ impulsive noise control\end{tabular}               &\begin{tabular}[c]{@{}l@{}}  \cite{sun2006adaptive} \end{tabular}\\ \hline
            2007       &\begin{tabular}[c]{@{}l@{}} Das, Panda,\\and Kuo \end{tabular}
            &\begin{tabular}[c]{@{}l@{}}Reduced-structure of fast Fourier transform (FFT)\\-based block
                filtered-x LMS and fast Hartley transform\\ (FHT)-based block filtered-x LMS algorithms \end{tabular}               &\begin{tabular}[c]{@{}l@{}} \cite{das2007new} \end{tabular}\\ \hline
            2007       &\begin{tabular}[c]{@{}l@{}} Sun and Kuo \end{tabular}
            &\begin{tabular}[c]{@{}l@{}}Cascade FxLMS algorithm for NANC \end{tabular}               &\begin{tabular}[c]{@{}l@{}}  \cite{sun2007active} \end{tabular}\\ \hline
            2007       &\begin{tabular}[c]{@{}l@{}} Carini and\\Sicuranza \end{tabular}
            &\begin{tabular}[c]{@{}l@{}}Optimal regularized FxAP algorithm \end{tabular}               &\begin{tabular}[c]{@{}l@{}} \cite{carini2007optimal} \end{tabular}\\ \hline
            2007       &\begin{tabular}[c]{@{}l@{}} Zhou and\\ DeBrunner \end{tabular}
            &\begin{tabular}[c]{@{}l@{}}FxLMS algorithm based on geometric analysis and the\\ strict positive real (SPR) property, and without\\ secondary path identification for single-tone noises\end{tabular}               &\begin{tabular}[c]{@{}l@{}} \cite{zhou2007new} \end{tabular}
           \\   \hline\hline
            \end{tabular}
        \label{Table001}
\end{table}

\subsection{Filtered-x ANC family}

Some fundamental methods of the filtered-x ANC family have been developed before the last decade. These contributions are summarized in Table \ref{Table001}. The papers listed in this review in some cases are extensions or variations of the fundamental methods.

\begin{figure}[h]
    \centering
    \includegraphics[scale=0.5] {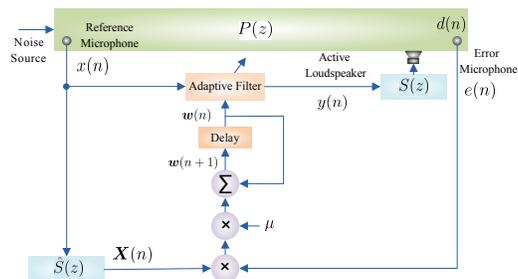}
    \caption{\label{03h} Diagram of feedforward ANC system with the FxLMS algorithm.}
    \label{FigFxLMS}
\end{figure}

\subsubsection{Filtered-x LMS-based algorithms}
The FxLMS algorithm has low computational complexity for linear ANC, and this algorithm is the cornerstone of many other algorithms. It can be used with feedforward, feedback, and hybrid ANC systems. The block diagram of the feedforward ANC system using the FxLMS-based algorithms is shown in Fig. \ref{FigFxLMS}. The residual noise (error signal) can be expressed as
\begin{equation}
e(n) \triangleq d(n) - s(n)*y(n)
\label{001}
\end{equation}
where $*$ denotes the convolution operation, and $s(n)$ stands for the impulse response of $S(z)$. The output of the controller is given by $y(n)=\bm w^{\mathrm T}(n)\bm x(n)$, where $\bm x(n) = [x(n),x(n-1),\ldots,x(n-M+1)]^{\mathrm T}$ is the input vector, $(\cdot)^{\mathrm T}$ is the transpose operation, $M$ denotes the filter length, and $\bm w(n)$ is the weight vector of the controller. The update equation of the basic FxLMS algorithm is expressed as \cite{morgan2013history}
\begin{equation}
\bm w(n+1) = \bm w(n) + \mu e(n)\bm X(n)
\label{002}
\end{equation}
where $\mu$ is the step size and $\bm X(n)=s(n)*\bm x(n)$ is the secondary signal. In particular, for global active control of noise inside a cavity, the modal FxLMS algorithm is adopted \cite{puri2018modal,puri2019global}. The conventional FxLMS algorithm is formulated in the modal domain of the acoustic cavity to reduce the specific acoustic modes. The modal FxLMS algorithm brings the concept of `modal secondary path' and `modal secondary signal'. Instead of the reference signal filtered using physical secondary paths, the modal FxLMS algorithm obtained $\bm X(n)$ by reference signal filtered employing modal secondary paths. By this way, the computational burden associated with filtering of the $\bm x(n)$ with $S(z)$ can be reduced, and acoustic potential energy can be reduced for the global noise control.

In the following, we summarize the development of the FxLMS-based algorithms from the type of undesired signal.

\noindent\textit{1) FxLMS-based algorithms for broadband noise}

\textit{$\bullet$ Theoretical analysis:} In order to ensure that the achievable reduction level predicted by theory can be realized in practice, effective theoretical analysis is crucial for the FxLMS algorithm. The analysis of the FxLMS algorithm has been deeply studied not only in the past decade, but also during the past few years. The development of FxLMS analysis before the past decade is summarized in Table \ref{FD}.

\begin{table}[h]
    \scriptsize
    \centering \doublerulesep=0.05pt
    \caption{Development of FxLMS analysis before the past decade.}
        \begin{tabular}{l|l|l|l|l}
            \hline
            \textbf{Years} & \textbf{Authors} & \textbf{Conditions} & \textbf{Contributions} & \textbf{References} \\ \hline
            1989 &\begin{tabular}[c]{@{}l@{}} Long, Ling,\\ and Proakis \end{tabular} &\begin{tabular}[c]{@{}l@{}} The secondary path is a delay system and the\\ input signal is a
broadband white signal\end{tabular} &\begin{tabular}[c]{@{}l@{}} Analysis of behavior of the\\ delayed LMS algorithm \end{tabular} &\cite{long1989lms,long1992corrections}\\ \hline
            1995 &\begin{tabular}[c]{@{}l@{}} Bjarnason \end{tabular} &\begin{tabular}[c]{@{}l@{}} The secondary path is a delay system and the\\ input signal is a Gaussian or colored Gaussian\\ signals\end{tabular} &\begin{tabular}[c]{@{}l@{}}Analysis of the FxLMS\\ algorithm with offline and \\online estimation of the\\ error-path filter \end{tabular} &\cite{bjarnason1995analysis}\\ \hline
            2000 &\begin{tabular}[c]{@{}l@{}} Tobias,\\ Bermudez,\\ and Bershad \end{tabular} &\begin{tabular}[c]{@{}l@{}} Imperfect secondary path estimation and the\\ input signal is a white or colored reference\\ signals\end{tabular} &\begin{tabular}[c]{@{}l@{}} Analysis of the FxLMS\\ algorithm without\\ independent assumption \end{tabular} &\cite{tobias2000mean}\\ \hline
            2007 &\begin{tabular}[c]{@{}l@{}} Fraanje et al. \end{tabular} &\begin{tabular}[c]{@{}l@{}} Asymptotically convergence of FxLMS \end{tabular} &\begin{tabular}[c]{@{}l@{}}Analysis of the robustness\\ of the FxLMS algorithm\end{tabular}&\cite{fraanje2007robustness,fraanje2007robustness2}\\ \hline
            2007 &\begin{tabular}[c]{@{}l@{}} Barrault,\\ Bermudez,\\ and Lenzi \end{tabular} &\begin{tabular}[c]{@{}l@{}} Performance of FxLMS in a finite duct \end{tabular} &\begin{tabular}[c]{@{}l@{}} Using a stochastic differential\\ equation (SDE) to analyze\\ the performance of the\\ FxLMS algorithm \end{tabular} &\cite{barrault2007new}\\ \hline
    \end{tabular}
    \label{FD}
\end{table}

In \cite{ardekani2010theoretical}, the FxLMS algorithm was analyzed based on the assumption that the secondary path is a moving average (MA) process and the stochastic input signal, which surmounts the limitation of deterministic input signal\footnote{The FxLMS in deterministic input signal was analyzed in  \cite{vicente2006novel}, see Table \ref{NNA}.}. Furthermore, in \cite{Iman2013Stochastic}, a new theoretical analysis of the FxLMS algorithm was conducted based on the method in \cite{ardekani2010theoretical} which considers the general secondary path cases. Aiming at revealing the convergence property of ANC systems with online secondary path estimation, the analysis of the broadband FxLMS algorithm was performed in \cite{chan2012performance}. In \cite{miyoshi2011statistical-mechanics}, an interesting trial was attempted to introduce the statistical-mechanics approach to analyze the FxLMS algorithm. According to this theory, the models and variables are represented by the cross-correlation between the elements or the auto-correlation of the elements. By making use of differential equations, the dynamical behaviors of the direction cosines among the vectors of an adaptive filter, the shifted filters, and an unknown system are described as the macroscopic variables. Such approach does not employ the independence assumption and small step size condition, which are widely used in the other studies. Follow-up works can be seen in \cite{miyoshi2015statistical,murata2017statistical}.

By assuming an exact secondary path model and the root locus analysis method, the behavior of the FxLMS-based ANC systems was investigated in \cite{ardekani2011stability,TabatabaeiRoot,ardekani2013stability}. By further adding the error of secondary path model in root locus analysis of the FxLMS algorithm, the effect of the secondary path model was clearly illustrated in \cite{ardekani2015efficient}. Analysis results showed that the FxLMS algorithm can be guaranteed to converge when $\hat S(z)$ and $S(z)$ have the same signs. Moreover, it confirms the existence of a simple secondary path model whose single non-zero coefficient can maintain the performance of the FxLMS algorithm \cite{ardekani2015efficient}.

\textit{$\bullet$ Leaky algorithm:} In certain scenarios, the conventional FxLMS algorithm may suffer from numerical problems and stagnation behavior due to inadequacy or low amplitude of the noise source \cite{AlFiltered}. In such a case, the leaky FxLMS (LFxLMS) algorithm and its variants were further developed within the last decade \cite{AlFiltered,cheer2016active}. The adaptation of the LFxLMS algorithm is given by \cite{cheer2016active}
\begin{equation}
\bm w(n+1) = (1-\mu\gamma)\bm w(n) + \mu e(n)\bm X(n),
\label{003}
\end{equation}
where $\gamma>0$ denotes the leakage factor. For $\gamma=0$, the LFxLMS algorithm reduces to the classical FxLMS algorithm. Recently, the optimal leaky factor was proposed based on the Karush-Kuhn-Tuker (KKT) condition for the output-constrained LFxLMS algorithm \cite{shi2019optimal}. Such approach provides an explicit criterion for selection of the leaky factor.

\textit{$\bullet$ HOEP criterion:} The high-order error moment (HOEP) can extract extra information from the signals as compared with the minimum mean square error (MMSE) criterion. Moreover, this criterion provides a more general framework for non-Gaussian signal processing. When the environment includes some non-Gaussian components, the HOEP can improve the filtering performance. Therefore, algorithms based on the HOEP criterion may be better than the MMSE-based algorithm, such as the FxLMS algorithm. By using the HOEP criterion, the filtered-x least mean kurtosis (FxLMK) algorithm was derived in \cite{lu2017improved}, which incorporates the least mean kurtosis (LMK) algorithm into feedforward ANC systems. Then, an improved version of the FxLMK algorithm was proposed, which can consistently estimate the parameters, without the need to acquire \textit{prior information} of the noise. Similar works can also be found in \cite{AlFiltered}, in which the least mean fourth (LMF) algorithm is integrated with a leaky strategy, resulting in the leaky filtered-x LMF (LFxLMF) algorithm. These methods achieve improvements in terms of stability.

\textit{$\bullet$ Frequency-domain algorithms:} When the sampling-frequency of the ANC system is high, the length of $\hat S(z)$ and $\bm w(n)$ will be large, leading to high computational complexity of the FxLMS algorithm. To reduce the computational complexity of ANC systems with high order, many frequency-domain ANC algorithms have been proposed. In these works and other similar references on the topic, the solutions generally rely on the use of a time-frequency domain wavelet packet \cite{luo2017a,padhi2018a,qiu2016multi}, Fourier transform \cite{rout2015computationally,tang2012time,zhang2019a,chan2012new,zecevic2015frequency}, and data block processing \cite{rout2015computationally,zhang2019a,zecevic2015frequency} for performance improvement. In \cite{zheng2013blind}, a new blind multi-channel ANC algorithm was proposed, and time and frequency-domain adaptive algorithms were developed. Blind pre-processing systems can pre-whiten the output as needed and as such time and frequency-domain adaptive algorithms converge faster than the basic multi-channel FxLMS algorithm. A piezoelectric feedback system with the discrete wavelet transform (DWT)-FxLMS algorithm was proposed to suppress noise inside vehicles \cite{wang2020active}. According to the structural vibration and acoustic characteristics of a simplified vehicle cavity model, it can be seen that such method has high application potential for active noise and vibration control (ANVC) systems.

\noindent\textit{2) FxLMS-based algorithms for narrowband noise}

The FxLMS-based algorithm can effectively tackle the noise from NANC systems. So far, a large number of FxLMS improvements have been proposed and analyzed for NANC, see \cite{kozacky2014convergence,kozacky2014cascaded,padhi2019new,huang2013a,xiaoseries,xiao2009properties,bo2014a,bo2014aa,matsuo2015on,mondal2018all-pass}. In Table \ref{NNA}, we summarize the development of FxLMS analysis in the context of NANC systems before the past decade. By supposing that the reference is synchronously-sampled,  the state-space representation was suggested for analyzing a general FxLMS algorithm \cite{haarnoja2013exact}. Moreover, the multi-channel version and its common narrowband modification were also covered by this form.  The performance analysis of the FxLMS algorithm in feedback ANC systems with internal model control (IMC) was also performed in \cite{Tongwei2014Stochastic} for band-limited white noise and it was subsequently improved by adaptive notch filtering (ANF) \cite{wang2014new}. Moreover, it overcame a difficulty that the feedback ANC system has a degraded performance for controlling broadband noise. The convergence properties of the FxLMS algorithm for traditional and new NANC systems have also been studied \cite{xiao2010new,matsuo2015on}.

\begin{table}[h]
    \scriptsize
    \centering \doublerulesep=0.05pt
    \caption{Development of FxLMS analysis for NANC systems before the past decade.}
        \begin{tabular}{l|l|l|l|l}
            \hline
            \textbf{Years} & \textbf{Authors} & \textbf{Conditions} & \textbf{Contributions} & \textbf{References} \\ \hline
            1999 &\begin{tabular}[c]{@{}l@{}} Bermudez and\\ Bershad \end{tabular} &\begin{tabular}[c]{@{}l@{}} The secondary path is linear time-invariant\\ filters and the input signal is a deterministic\\ sinusoid signal\end{tabular} &\begin{tabular}[c]{@{}l@{}}Non-Wiener behavior of the\\ filtered LMS algorithm\end{tabular}  & \cite{bermudez1999non} \\ \hline
            2006 &\begin{tabular}[c]{@{}l@{}} Vicente and\\ Masgrau \end{tabular} &\begin{tabular}[c]{@{}l@{}} The secondary path is a delay system and the\\ input signal is a narrowband signal\end{tabular} &\begin{tabular}[c]{@{}l@{}}FxLMS convergence condition\\
with deterministic reference\end{tabular}  & \cite{vicente2006novel} \\ \hline
            2008 &\begin{tabular}[c]{@{}l@{}} Xiao \end{tabular} &\begin{tabular}[c]{@{}l@{}} The secondary path is modeled by MA process\\ and the input signal is a narrowband signal\end{tabular} &\begin{tabular}[c]{@{}l@{}}Stochastic analysis of the\\ FxLMS for NANC\end{tabular} & \cite{xiao2008stochastic} \\ \hline
    \end{tabular}
    \label{NNA}
\end{table}

In \cite{chang2013complete,chang2013complete2}, the FxLMS algorithm was employed by a parallel or direct NANC system. Moreover, the corresponding convergence analyses for parallel NANC and direct/parallel NANC systems were conducted. All the results demonstrated that the parallel or direct/parallel NANC can converge faster than the conventional NANC.

It is well known that the fixed step size can lead to a trade-off between convergence rate and noise residue. Therefore, it is natural to utilize variable step size (VSS) schemes in the conventional FxLMS algorithm. As for the NANC system, a VSS-FxLMS algorithm that has fast convergence, improved tracking capability, and small noise residue, was proposed \cite{huang2013a}. Simulations demonstrated that the VSS-FxLMS algorithm even achieved ameliorated performance than the FxRLS algorithm in non-stationary noise environments \cite{huang2013a}. Some similar alternating VSS approaches have been proposed for secondary path identification \cite{narasimhan2010variable} and for NANC systems \cite{xiao2011new,delega2017a}.

\noindent\textit{3) FxLMS-based algorithms for impulsive noise}

The impulsive noise is often due to the occurrence of noise disturbance with low probability but large amplitude, which has become a great challenge for ANC systems \cite{akhtar2009improving}. The $\alpha$-stable noise can effectively model the impulsive noise encountered in ANC systems, which explains why such noise is widely used for active impulsive noise control (AINC) \cite{akhtar2009improving,zhou2015active,wu2013active}.

To combat $\alpha$-stable noise, a clipped FxLMS algorithm was proposed by Sun et al., which puts limitations on the input signal \cite{sun2006adaptive}. In 2009, the improved version of Sun's algorithm was proposed and termed as Akhtar's algorithm \cite{akhtar2009improving}. The Akhtar's algorithm has a restriction in both input signal and error signal, which can be expressed as:
\begin{subequations}
\begin{equation}
\bm w(n+1) = \bm w(n) + \mu e'(n)(s(n)*\bm x'(n))
\label{004a}
\end{equation}
where
\begin{equation}
e'(n)=\left\{\begin{array}{l}
c_1,\;\;\;\;\;\;{\mathrm {if}}\;e(n) \leq c_1 \\
c_2,\;\;\;\;\;\;{\mathrm {if}}\;e(n) \geq c_2 \\
e(n),\;\;\;{\mathrm {otherwise}}
\end{array} \right.
\label{004b}
\end{equation}
and
\begin{equation}
x'(n)=\left\{\begin{array}{l}
c_1,\;\;\;\;\;\;{\mathrm {if}}\;x(n) \leq c_1 \\
c_2,\;\;\;\;\;\;{\mathrm {if}}\;x(n) \geq c_2 \\
x(n),\;\;\;{\mathrm {otherwise}}
\end{array} \right.
\label{004c}
\end{equation}
\end{subequations}
where $c_1>0$ and $c_2>0$ are two threshold parameters.

To further enhance the performance, an FxlogLMS algorithm, which minimizes the squared logarithmic transformation of the error signal, was proposed in \cite{wu2011an}. The update equation of the FxlogLMS algorithm is described by
\begin{equation}
\bm w(n+1) = \bm w(n) + \mu {\mathrm {sgn}}\{e(n)\}\frac{\log|e(n)|}{|e(n)|}\bm X(n)
\label{005}
\end{equation}
and for $|e(n)|<1$, setting $|e(n)|=1$. In this expression, ${\mathrm {sgn}}\{\cdot\}$ denotes the sign function.

A computationally fast algorithm that uses the \textit{binormalized data reusing} was proposed in \cite{akhtar2016binormalized}, based on the FxLMS algorithm. Related to the binormalized data reusing is the \textit{data reusing} method \cite{akhtar2015data-reusing-based}. The filtered-x data reusing algorithm solves the same optimization problem associated with the classical AP approaches, but employs an iterative strategy to define the projection onto a set of hyperplanes instead of using past information directly.

Based on the fractional lower order moment (FLOM) criterion, a class of the filtered-x least mean $p$th power (FxLMP) algorithms was investigated for robust performance in the presence of $\alpha$-stable noise \cite{akhtar2015data-reusing-based,akhtar2019a,MuhammadImproving}. A representative algorithm of this type is the modified normalized FxLMP (MNFxLMP) algorithm \cite{MuhammadImproving}. Furthermore, a filtered-x general step size NLMS (FxgsnLMS) algorithm \cite{zhou2015active} and a companding FxsgnLMS algorithm \cite{tan2015active} were developed. The essence of the former is the adaptation of Gaussian kernel, and the latter is based on the instantaneous power of the companded error signal. In \cite{bergamasco2012active}, an online estimation approach for non-Gaussian noise characteristics was developed and then was incorporated into the sign FxLMS and FxLMP algorithms. Such algorithm avoids the selection problem of threshold parameters $c_1$ and $c_2$, and it is easy to implement.

The Akhtar's algorithm and its variants utilized the hard limit to clip the residual noise and the reference signal \cite{mirza2019,meng2020modified}. Accordingly, the algorithm in \cite{saravanan2019an} originally introduced a soft bound for residual noise and reference signal, which offers lower noise reduction level for impulsive noise. Recalling the M-estimation has robustness for system identification, several variants of the M-estimate algorithm were also presented for AINC \cite{wu2013an,li2013active,sun2015enhanced,sun2015a}.

\begin{figure}[!htb]
    \centering
    \includegraphics[scale=0.5] {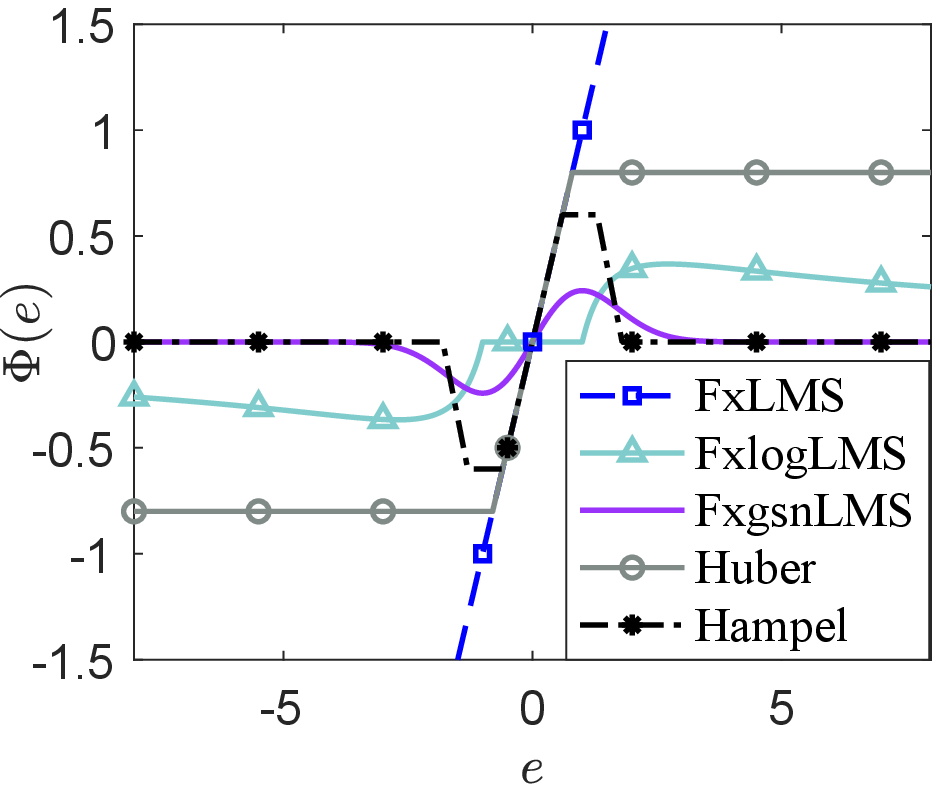}
    \caption{\label{04} Score function of the FxLMS algorithm \cite{morgan2013history}, FxlogLMS algorithm \cite{wu2011an}, FxsgnLMS algorithm \cite{zhou2015active}, Huber's-based FxLMS algorithm \cite{wu2013an}, and Hample's-based FxLMS algorithm \cite{wu2013an}.}
    \label{Fig04}
    \includegraphics[scale=0.5] {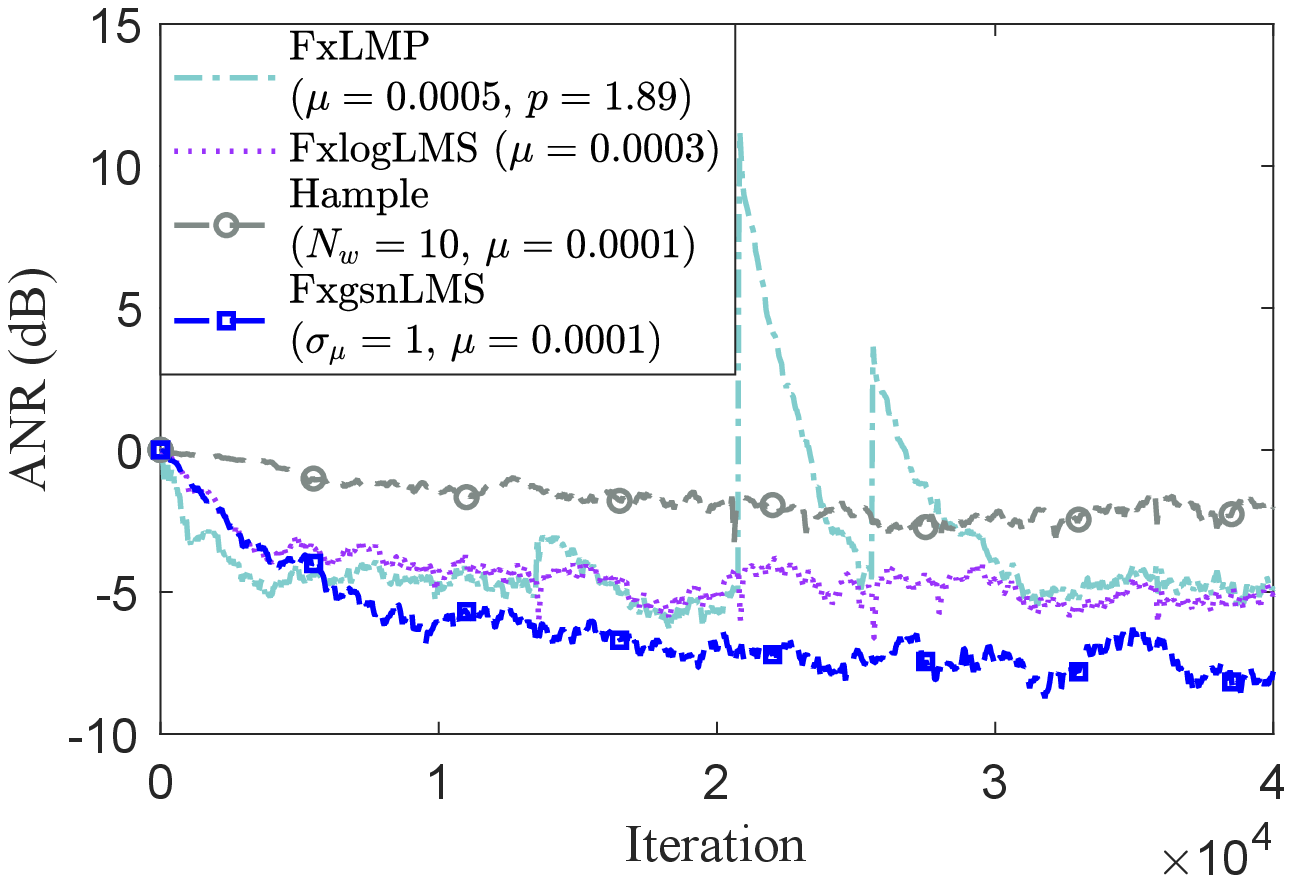}
    \caption{ANRs of the representative AINC algorithms in $\alpha=1.9$, where the simulation settings are the same as \cite{lu2017improved}.}
    \label{Simu1}
\end{figure}

Let us define $J(n)$ as the cost function and $\Phi(e)=\partial J(e)/ \partial e$ as the score function. Fig. \ref{Fig04} shows the score function $\Phi(e)$ in the FxLMS-based algorithms. As can be seen, the score function of the FxLMS algorithm is unbounded. In contrast, the M-estimator can bound the outliers from impulsive noise. The algorithm in \cite{sun2015a} used a family of robust estimators, such as \textit{Huber}, \textit{Fair}, and \textit{Hample} for combating impulsive noise, and further utilized the threshold scheme from Akhtar's algorithm. For a fair comparison, the averaged noise reduction (ANR) is usually employed as a performance measure, which is defined by \cite{akhtar2009improving}
\begin{equation}
\mathrm{ANR}(n) \triangleq 20\log\left\{ \frac{A_e(n)}{A_d(n)} \right\}
\label{s1}
\end{equation}
where $A_e(n) = \chi A_e(n-1)+(1-\chi)|e(n)|$ and $A_d(n) = \chi A_d(n-1)+(1-\chi)|d(n)|$, and $\chi=0.999$. In Fig. \ref{Simu1}, we compare the ANRs of the representative algorithms. The primary path $P(z)$ and the secondary path $S(z)$ are modeled by an FIR filter with length 256 and 100, respectively. The filter length is set to 128 \cite{lu2017improved}. With similar convergence rate, the FxgsnLMS algorithm has the smallest noise residual in this scenario.

\subsubsection{Filtered-x AP-based algorithms}

\noindent\textit{1) FxAP-based algorithms for broadband noise}

The AP algorithm updates the weights on the basis of multiple input vectors to accelerate convergence speed if driven by highly correlated input signals. For these reasons, it has become a good alternative to LMS-type controllers. The update equation for the basic FxAP algorithm is \cite{ferrer2012steady-state}
\begin{equation}
\bm w(n+1) = \bm w(n) + \mu \bm U(n)\left[\bm U^{\mathrm {T}}(n)\bm U(n)\right]^{-1}\bm e(n)
\label{008}
\end{equation}
where $\bm e(n)=[e(n),e(n-1),\ldots,e(n-P+1)]^{\mathrm {T}}$, $\bm U(n)=[\bm x(n),\bm x(n-1),\ldots,\bm x(n-P+1)]$ is a $M \times P$ matrix, and $P$ denotes the projection order.

In \cite{ferrer2011transient,ferrer2012steady-state}, the transient and steady-state performance of the FxAP algorithm was analyzed for multi-channel ANC systems, which is based on the energy conservation argument and does not require a specific characteristic of the signal. The optimal VSS-FxAP algorithm was developed by performing analysis of mean square deviation (MSD) and its non-stationary version was also proposed \cite{song2015an}. A more accurate result was presented in \cite{guo2020convergence,guomean}, which conducts the augmented weight-error vector for convergence analysis.

\noindent\textit{2) FxAP-based algorithms for impulsive noise}

The above mentioned algorithms are also challenged by the difficulty of converging in the presence of impulsive noise. A simple, yet effective approach for AINC was proposed in \cite{xiao2016a}, which is derived by introducing the efficient AP sign (APS) algorithm into ANC systems, resulting in the FxAPS algorithm. Additionally, two extensions regarding a VSS scheme and partial update (PU) were developed.

\subsubsection{Filtered-x RLS-based algorithms}

\noindent\textit{1) FxRLS-based algorithms for broadband noise}

The standard FxRLS algorithm can converge faster than the FxLMS algorithm, at the price of increased complexity \cite{reddy2011hybrid}. As such, the RLS algorithm has been extended to ANC systems in recent years. In \cite{reddy2011hybrid}, a hybrid algorithm, which can switch between the filtered-x NLMS (FxNLMS) and FxRLS algorithms, was proposed for functional magnetic resonance imaging (fMRI) acoustic noise control. The FxRLS algorithm is adopted at the initial convergence stage for obtaining fast convergence. Once detected it stops converging, the hybrid algorithm switches to the FxNLMS algorithm for low residual error. By doing this, the overall attenuation performance can be better than that of the individual FxRLS and FxNLMS components.

\noindent\textit{2) FxRLS-based algorithms for narrowband noise}

There is scarce literature focused on using the FxRLS-type algorithm for NANC. In \cite{aslam2019robust}, a filtered-x optimally weighted RLS (FxOWRLS) algorithm was derived for both feedforward and feedback ANC systems with bounded narrowband disturbances, which can reduce the computational burden by reducing the update ratio in the adaptation process.

\noindent\textit{3) FxRLS-based algorithms for impulsive noise}

In \cite{wu2015a}, to suppress the effect of the impulsive noise, a logarithmic cost function has been employed with the consideration of \textit{communication error} in the error signal. Following this work, several FxRLS variants have been developed for AINC \cite{mirza2016robust,lu2017active}. In \cite{mirza2016robust}, a modified FxRLS algorithm was proposed by addition of state-space for AINC. The filtered-x recursive maximum correntropy (FxRMC) algorithm \cite{lu2017active} presented an information theoretic learning (ITL) approach for AINC. Moreover, an adaptive kernel size scheme was introduced. Note that such maximum correntropy criterion (MCC) can also be utilized in NLANC systems, see Part II of this work. Another FxRLS-based algorithm for AINC was proposed in \cite{zeb2017improving}, where Akhtar's scheme \cite{akhtar2009improving} was incorporated with the modified gain method.

\subsubsection{Subband ANC algorithms}

\noindent\textit{1) Subband ANC algorithms for broadband noise}

\begin{figure}[!htb]
    \centering
    \includegraphics[scale=0.5] {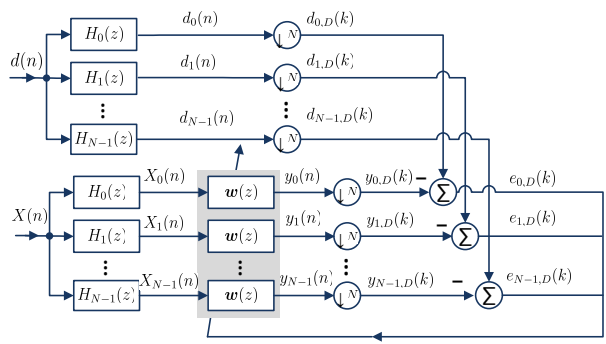}
    \caption{\label{10}Diagram of the multi-band structured SAF.}
    \label{Fig10}
\end{figure}
\begin{figure}[!htb]
    \centering
    \includegraphics[scale=0.45] {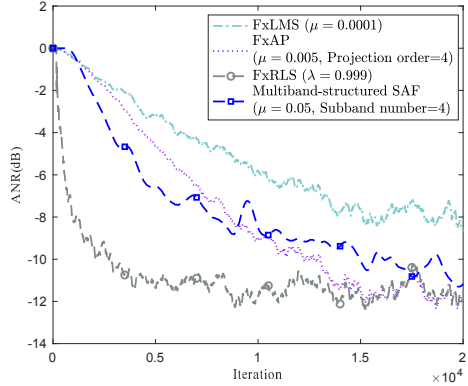}
    \caption{ANRs of the algorithms with $\alpha=2$, where the simulation settings are the same as \cite{lu2017improved}.}
    \label{Simu2}
\end{figure}

To deal with the long channel responses and colored inputs in ANC systems, the subband adaptive filter (SAF) was developed for fast convergence and low computational complexity \cite{chu2015a}. A typical SAF based on multi-band structure with $N$ subbands can be seen in Fig. \ref{Fig10}. The signal $d(n)$ and $X(n)$ are decomposed through the analysis filters $H_i(z)$, $i=0,\ldots,N-1$. The subband reference signals $X_i(k)$ are filtered by the adaptive filter to generate the subband output signals $y_i(n)$. Then, the subband signals $d_i(n)$ and $y_i(n)$ are critically decimated to lower sampled rate sequences $d_{i,D}(k) = d_i(kN)$ and $y_{i,D}(k)= y_i(kN)$, where $n$ and $k$ are used to index the original sequences and the decimated sequences. The factor $D$ denotes the decimation factor, which is chosen as same as the number of the subband filters to prevent aliasing of the signals. Finally, the decimated error signal $e_{i,D}(k) = d_{i,D}(k)-y_{i,D}(k)$ is utilized for adjusting of the subband ANC controller.

To assess the performance, in Fig. \ref{Simu2}, the ANRs of the LMS and non-LMS-based algorithms are investigated. In this case, the primary path $P(z)$ and the secondary path $S(z)$ are generated by FIR filter with length of 256 and 100, respectively \cite{lu2017improved}. The length of the adaptive filter is set to 128. The $\alpha$-stable noise with $\alpha=2$ is adopted as the noise source, which corresponds to the Gaussian distribution. One can observe that the FxRLS algorithm has the fastest convergence rate and the FxLMS algorithm suffers from slow convergence.
Then, a delayless SAF algorithm for multi-input multi-output (MIMO) ANC applications was developed, which is based on Milani's work \cite{milani2009a,milani2010analysis} and can significantly mitigate the computational cost \cite{cheer2017an}.

\noindent\textit{2) Subband ANC algorithms for narrowband noise}

For most ANC systems, it is necessary to estimate the secondary path offline or online, which undoubtedly increases the computational complexity. Therefore, some subband ANC algorithms consider avoiding the secondary path estimation to reduce complexity. However, when the secondary path phase is close to $\pm 90^{\mathrm o}$, the convergence rate of the algorithm is slow. To overcome this limitation, a frequency-domain delayless subband algorithm was proposed, where 4 update directions, $180^{\mathrm o}$, $0^{\mathrm o}$, and $\pm 90^{\mathrm o}$ are adopted to single-tone and narrowband noise control \cite{wu2008improved}. The disadvantage of this method is that the filter update process must be performed in the frequency-domain, resulting in high computational complexity. To overcome this limitation, in \cite{gao2016simplified}, a simplified attempt was developed to cope with the implementation problem when the secondary path phase is close to $\pm 90^{\mathrm o}$. Two reference signals are generated in each
subband, and two update directions are utilized. Only one subband reference signal and one update direction are employed to approximate the phase response of the residual secondary path. Then, the coefficients of the full-band adaptive controller are directly adapted in time-domain.

\noindent\textit{3) Subband ANC algorithms for impulsive noise}

The above mentioned SAF algorithm may fail to work in the presence of impulsive noise since the adaptation is based on the MMSE criterion. To fill this gap, a VSS normalized SAF (VSS-NSAF) was introduced for ANC system, whose step size is adapted to prevent the wrong update by impulsive noise \cite{park2018a}. Since impulsive noise is a great challenge for ANC systems, we summarize the above mentioned contributions of AINC In Table \ref{Table02}.

\begin{table}[htp]
    \scriptsize
    \centering \doublerulesep=0.05pt
    \caption{Contributions of AINC in the past decade.}
    \begin{tabular}{l|l|l|l}
        \hline \hline
        \begin{tabular}[c]{@{}l@{}}\textbf{Akhtar's algorithm}\\ \textbf{and its variants}\end{tabular} &
        \begin{tabular}[c]{@{}l@{}}\textbf{FxlogLMS}\end{tabular} &
        \begin{tabular}[c]{@{}l@{}}\textbf{HOEP/FLOM based}\\ \textbf{algorithms}\end{tabular} &
        \begin{tabular}[c]{@{}l@{}}\textbf{Soft bound}\\ \textbf{algorithms}\end{tabular} \\ \hline
        \begin{tabular}[c]{@{}l@{}} \cite{akhtar2009improving,mirza2019}  \\ \cite{sun2015a,meng2020modified} \end{tabular}  &
        \cite{wu2011an} &
        \begin{tabular}[c]{@{}l@{}} \cite{lu2017improved,AlFiltered,zhou2015active} \\
            \cite{akhtar2015data-reusing-based,akhtar2019a,MuhammadImproving,tan2015active,bergamasco2012active,mirza2019}\end{tabular} &
        \cite{saravanan2019an} \\ \hline \hline
        \begin{tabular}[c]{@{}l@{}}\textbf{M-estimate based}\\ \textbf{algorithms}\end{tabular} &
        \begin{tabular}[c]{@{}l@{}}\textbf{FxAP-based}\\ \textbf{algorithms}\end{tabular} &
        \begin{tabular}[c]{@{}l@{}}\textbf{FxRLS-based}\\ \textbf{algorithms}\end{tabular} &
        \begin{tabular}[c]{@{}l@{}}\textbf{SAF-based}\\ \textbf{algorithms}\end{tabular} \\ \hline
        \cite{wu2013an,li2013active,sun2015enhanced,sun2015a} &
        \cite{xiao2016a} &
        \begin{tabular}[c]{@{}l@{}} \cite{wu2015a,mirza2016robust}  \\  \cite{lu2017active,zeb2017improving} \end{tabular}  &
        \cite{park2018a} \\ \hline \hline
    \end{tabular}
    \label{Table02}
\end{table}

\subsubsection{Lattice ANC algorithms}

The lattice filter is also an important architecture in ANC systems. Such structure can attenuate multiple sinusoidal interferences in ANC systems and its corresponding algorithm, i.e., the gradient adaptive lattice (GAL) algorithm, can provide a reliable performance as compared with the known algorithm \cite{chen2015improving,kim2012a}. In the past decade, two VSS strategies have been suggested for the GAL algorithm, resulting in two VSS filtered-x GAL (VSS-FxGAL) algorithms \cite{kim2012a,kim2013a}. These algorithms exhibit good attenuation performance for hybrid narrowband and broadband noise. Very recently, a recursive least-squares lattice (RLSL) algorithm grouping the secondary path innovation (SPI) and lattice-order decision (LOD) was developed  \cite{kim2020recursive}. The SPI algorithm whitens the error signal into a virtual error signal just before the secondary path to generate virtual undesired signals corresponding to the output of the lattice filter. The LOD algorithm determines the order of the lattice filter, while considering the noise reduction performance. As a consequence, a faster convergence rate and lower computational complexity is achieved as compared to the FxRLS algorithm.

\subsection{Filtered-e ANC family}

\begin{figure}[!htb]
    \centering
    \includegraphics[scale=0.5] {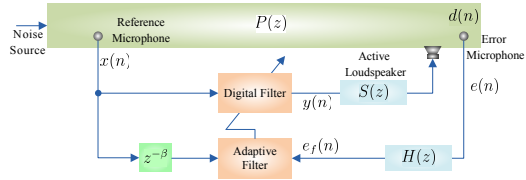}
    \caption{\label{05} Diagram of the FeLMS algorithm.}
    \label{Fig05}
\end{figure}
\begin{figure}[!htb]
    \centering
    \includegraphics[scale=0.5] {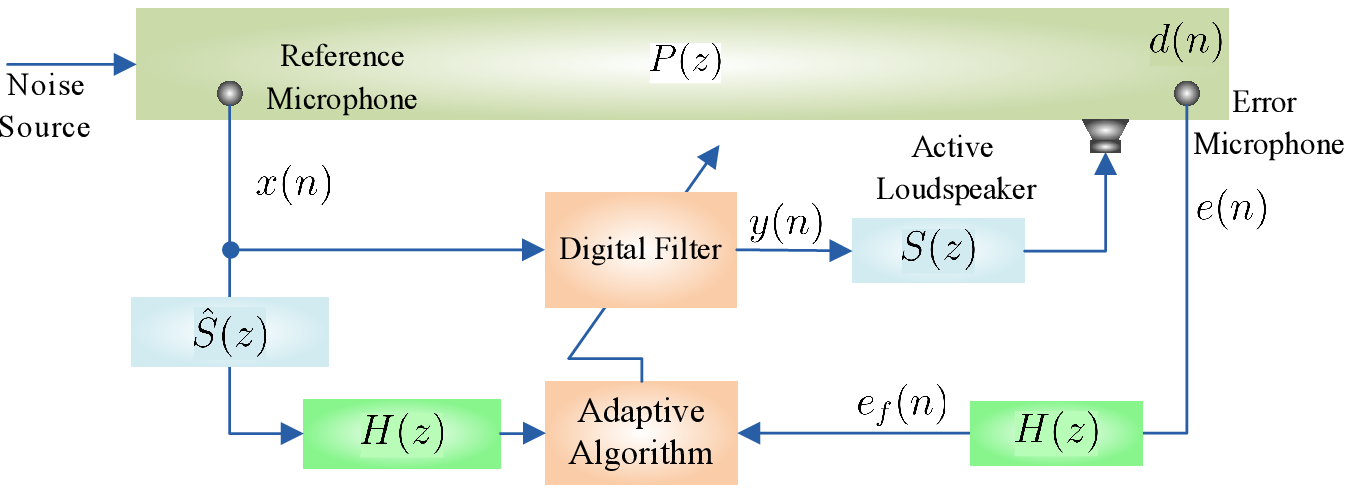}
    \caption{\label{06} Diagram of the FxFeLMS algorithm.}
    \label{Fig06}
\end{figure}

    \begin{table}[]
    \scriptsize
    \centering \doublerulesep=0.05pt
    \caption{Time line of filtered-e ANC algorithms.}
        \begin{tabular}{l|l|l|l}
            \hline
            \textbf{Years} & \textbf{Authors} & \textbf{Contributions} & \textbf{References} \\ \hline
1996       &\begin{tabular}[c]{@{}l@{}} Wan \end{tabular}         &\begin{tabular}[c]{@{}l@{}}Adjoint LMS algorithm\end{tabular}             &\begin{tabular}[c]{@{}l@{}} \cite{wan1996adjoint}\end{tabular}            \\ \hline
1996       &\begin{tabular}[c]{@{}l@{}} Rupp and Sayed \end{tabular}         &\begin{tabular}[c]{@{}l@{}}Analysis of the robustness of the FxFeLMS\\ algorithm along the line of $H_\infty$ theory\end{tabular}             &\begin{tabular}[c]{@{}l@{}} \cite{rupp1996time}\end{tabular}            \\ \hline
1999       &Sujbert         &\begin{tabular}[c]{@{}l@{}}FxFeLMS algorithm\end{tabular}               &\begin{tabular}[c]{@{}l@{}} \cite{sujbert1999new} \end{tabular}           \\ \hline
2005 &Miyagi and Sakai   &\begin{tabular}[c]{@{}l@{}}Analysis of the mean-square performance\\ of the FxFeLMS algorithm \end{tabular}  &\begin{tabular}[c]{@{}l@{}} \cite{miyagi2005mean}\end{tabular} \\ \hline
2006       &\begin{tabular}[c]{@{}l@{}} DeBrunner\\ and Zhou \end{tabular}
&\begin{tabular}[c]{@{}l@{}}Hybrid FeLMS algorithm for performance\\ improvement\end{tabular}               &\begin{tabular}[c]{@{}l@{}} \cite{debrunner2006hybrid} \end{tabular}             \\      \hline
\end{tabular}
\label{Table002}
\end{table}

The key contributions of the filtered-e ANC algorithms before the last decade are listed in Table \ref{Table002}. Note that the basic filtered-e LMS (FeLMS) and filtered-x FeLMS (FxFeLMS) algorithms have been developed in 1996 and 1999, respectively.

\subsubsection{Filtered-e LMS-based algorithms}
\noindent\textit{1) FeLMS-based algorithms for broadband noise}

The FxLMS algorithm may converge to a biased solution when the feedforward ANC system is corrupted by noise. Moreover, it is well known that the FxLMS algorithm suffers from slow convergence speed owing to the effect of $S(z)$. If $|S(z)|$ has high dynamics, there are certain frequency bands with small loop gain. As a result, the convergence rate for any signal appearing in this frequency range is slow.

To overcome these drawbacks, the FeLMS algorithm employs the error signal filtered by the error filter $H(z)$, instead of filtering the reference signals by $S(z)$. Two types of methods can be used to design $H(z)$. The first type of the FeLMS algorithm, termed as the adjoint LMS (ALMS) algorithm, has similar properties to the FxLMS algorithm. For the ALMS algorithm, $H(z)$ is designed as
\begin{equation}
    H(z) = z^{-\beta}\hat S^*(z)
    \label{bb1}
\end{equation}
where $\hat S^*(z)$ represents the adjoint transfer function of the secondary path and $\beta$ represents the number of delays.

The second type of the FeLMS algorithm, called the secondary path equalization (SPE) algorithm, has the same block diagram as the ALMS algorithm shown in Fig. \ref{Fig05}, where $e_f(n)$ is the residual noise $e(n)$ that has been filtered by the error filter $H(z)$. In this algorithm, $H(z)$ is designed as the (pseudo-)inverse of the secondary path filter
\begin{equation}
H(z) = \left[ z^{-\beta}\hat S^{-1}(z) \right]_+
\label{bb2}
\end{equation}
where $\hat S^{-1}(z) $ is the inverse transfer function of $\hat S(z)$, and $\left[ V(z) \right]_+$ stands for the casual part of $V(z)$. The weight update equation of the FeLMS algorithm is given by
\begin{equation}
\bm w(n+1) = \bm w(n) + \mu e_f(n)\bm X(n-\beta)
\label{007}
\end{equation}

To further accelerate the convergence rate of the FeLMS algorithm, the FxFeLMS algorithm was proposed, which can overcome the slow convergence of the FxLMS algorithm when the reference input is the sinusoidal signal \cite{sujbert1999new}. The structure of the FxFeLMS algorithm is plotted in Fig. \ref{Fig06}, where $H(z)$ is required in both input and error signal path. If $H(z)=1$, the FxFeLMS algorithm degenerates into the basic FxLMS algorithm. Under ideal situations, i.e., $S(z)=\hat S(z)$, $|H(z)|^2=1/|S(z)|^2$. In \cite{lopez2009modified}, a modified FxFeLMS algorithm was proposed, which is derived based on the stochastic model for the first and second moments of the FxFeLMS algorithm without the independent assumption.

\noindent\textit{2) FeLMS-based algorithms for narrowband noise}

The performance of the FxLMS-based algorithms for NANC systems is subjected to the number of targeted frequencies and the estimated secondary path. Therefore, it is natural to consider using FeLMS-based algorithms for NANC. In \cite{zhu2019a}, the FeLMS algorithm was intentionally introduced to deal with the noise in NANC systems. Moreover, the convergence behavior of the algorithm was analyzed.  This work has proven that FeLMS-type algorithms are feasible in dealing with narrowband noise. The existing works of NANC are outlined in Table \ref{NA}.

\begin{table}[h]
    \scriptsize
    \centering \doublerulesep=0.05pt
    \caption{Contributions of NANC in the past decade.}
   \begin{tabular}{l|l|l|l}
            \hline
            \begin{tabular}[c]{@{}l@{}}\textbf{FxLMS-based}\\ \textbf{algorithms}\end{tabular}&
            \begin{tabular}[c]{@{}l@{}}\textbf{Subband ANC}\\ \textbf{algorithms}\end{tabular}& \begin{tabular}[c]{@{}l@{}}\textbf{FxRLS-based}\\ \textbf{algorithms}\end{tabular} & \begin{tabular}[c]{@{}l@{}}\textbf{FeLMS-based}\\ \textbf{algorithms}\end{tabular} \\ \hline
            \cite{wu2014a,padhi2019new,huang2013a,xiaoseries,xiao2009properties,bo2014a,bo2014aa,matsuo2015on,mondal2018all-pass,xiao2010new,narasimhan2010variable,xiao2011new,delega2017a,chang2013complete,chang2013complete2,haarnoja2013exact,kozacky2014convergence,kozacky2014cascaded}                                                                                           & \multirow{3}{*}{\cite{gao2016simplified}(Also for}                                                & \multirow{3}{*}{}                                                                                                \\ \cline{1-1}
            \begin{tabular}[c]{@{}l@{}}\textbf{For uncorrelated narrowband disturbances}\end{tabular}           &                                                                  &\cite{aslam2019robust}(Also for                                                                                                                                   \\ \cline{1-1}
            \cite{akhtar2011improving,wu2015decoupling,padhi2019cascading}
            &band-limited                                                                                    &multi-tonal                                                                   &\cite{zhu2019a}                                                                                                                                   \\ \cline{1-1}
            \begin{tabular}[c]{@{}l@{}}\textbf{For band-limited white noises and}  \textbf{tonal signals}\end{tabular} &white noise)                                                                  &signals)                                                                                                                                    \\ \cline{1-1}
            \cite{wu2018a,Tongwei2014Stochastic,wang2014new}                                                                                           &                                                                  &                                                                                                                                    \\ \hline
    \end{tabular}
    \label{NA}
\end{table}

\subsection{Filtered-u ANC family}

    \begin{table}[h]
    \scriptsize
    \centering \doublerulesep=0.05pt
    \caption{Time line of filtered-u ANC algorithms.}
        \begin{tabular}{l|l|l|l}
            \hline
            \textbf{Years} & \textbf{Authors} & \textbf{Contributions} & \textbf{References} \\ \hline
            1987      &\begin{tabular}[c]{@{}l@{}} Eriksson, Allie,\\ and Greiner \end{tabular}
            &\begin{tabular}[c]{@{}l@{}}IIR  adaptive algorithm (recursive least mean square)\\ for ANC \end{tabular}               &\begin{tabular}[c]{@{}l@{}}  \cite{eriksson1987selection} \end{tabular}             \\      \hline
            1991      &\begin{tabular}[c]{@{}l@{}} Eriksson \end{tabular}
            &\begin{tabular}[c]{@{}l@{}}FuLMS algorithm \end{tabular}               &\begin{tabular}[c]{@{}l@{}}  \cite{eriksson1991development} \end{tabular}             \\      \hline
            1994       &\begin{tabular}[c]{@{}l@{}} Snyder \end{tabular}
            &\begin{tabular}[c]{@{}l@{}}FuLMS algorithm with simple hyper-stable adaptive\\
                recursive filter \end{tabular}               &\begin{tabular}[c]{@{}l@{}}  \cite{snyder1994active} \end{tabular}             \\      \hline
            2003       &\begin{tabular}[c]{@{}l@{}} Fraanje, Verhaegen,\\ and Doelman \end{tabular}
            &\begin{tabular}[c]{@{}l@{}}Analysis of the FuLMS algorithm and develop a\\ preconditioned FuLMS algorithm \end{tabular}               &\begin{tabular}[c]{@{}l@{}}  \cite{fraanje2003convergence} \end{tabular}             \\      \hline
            2003       &\begin{tabular}[c]{@{}l@{}} Lu et al. \end{tabular}
            &\begin{tabular}[c]{@{}l@{}} IIR filter with lattice form for ANC\end{tabular}               &\begin{tabular}[c]{@{}l@{}}  \cite{lu2003lattice} \end{tabular}  \\  \hline
            2004       &\begin{tabular}[c]{@{}l@{}} Sun and Meng\end{tabular}
            &\begin{tabular}[c]{@{}l@{}} Steiglitz-Mcbride type adaptive IIR algorithm for ANC \end{tabular}               &\begin{tabular}[c]{@{}l@{}} \cite{sun2004steiglitz} \end{tabular}  \\  \hline
    \end{tabular}
    \label{Table003}
\end{table}

Several fundamental filtered-u ANC algorithms have been proposed before the past decade. We summarize these efforts in Table \ref{Table003}. In the following, the development of filtered-u ANC algorithms in the past decade is reviewed in detail.

\subsubsection{Filtered-u LMS-based algorithms}
\noindent\textit{1) FuLMS-based algorithms for broadband noise}

\begin{figure}[!htb]
    \centering
    \includegraphics[scale=0.5] {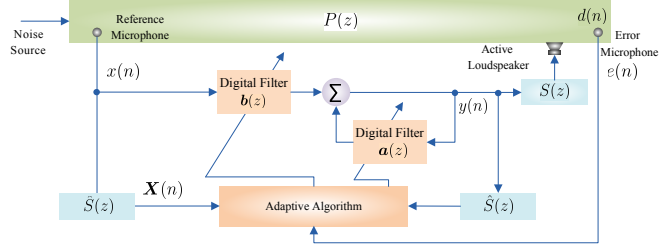}
    \caption{\label{07} Diagram of the FuLMS algorithm, where the zero and pole coefficients are updated by using one adaptive algorithm. The coefficients can be also independently updated by using two adaptive algorithms.}
    \label{Fig07}
\end{figure}

It is well known that the ANC algorithms can be extended to active vibration control (AVC) problem \cite{landau2013iir}. Among the ANC algorithms, the FuLMS algorithm can be effectively applicable to both AVC and ANC systems. The FuLMS algorithm is often used to update the weight vector of the IIR filter, whereas the above mentioned FxLMS and FeLMS algorithms are commonly adopted to adapt the coefficients of the FIR filter. Fig. \ref{Fig07} depicts the diagram of an ANC system using FuLMS algorithm. The iterative process of the pole and
zero-coefficients for the basic FuLMS algorithm can be summarized as follows \cite{kim2011modified-filtered-u}:
\begin{subequations}
\begin{equation}
\bm a(n+1) = \bm a(n) + \mu_1 e(n)\bm y(n-1)
\end{equation}
\begin{equation}
\bm b(n+1) = \bm b(n) + \mu_2 e(n)\bm X(n)
\label{007ab}
\end{equation}
\end{subequations}
where $\bm a(n)$ and $\bm b(n)$ are the weight vectors of poles and zeroes in the IIR filter at time $n$, respectively, $\mu_1$ and $\mu_2$ are the step sizes, and $\bm y(n-1)$ is the previous output vector. The FuLMS algorithm consumes fewer coefficients than the FIR filter and is particularly well-suited for ANC systems with short acoustic ducts \cite{kim2011modified-filtered-u,park2012a}. For example, in an automotive engine, the intake system owns a short duct and as such, the FuLMS-based algorithms are usually preferred. Unfortunately, the FuLMS algorithm cannot ensure global convergence because of the multimodal error surface and inherent limitation of the IIR filter. To solve the stability problem, new FuLMS algorithms were developed in \cite{kim2011modified-filtered-u,ho2020development}. In \cite{kim2011modified-filtered-u}, a modified FuLMS algorithm was developed by combining a simple hyper-stable adaptive recursive filter (SHARF) and a VSS scheme. By using the equation-error (EE), the FuLMS algorithm was derived according to the output-error (OE) model and its step size bound and global minimum were also analyzed \cite{ho2020development}.

\subsubsection{Filtered-u RLS-based algorithms}
\noindent\textit{1) FuRLS-based algorithms for broadband noise}

To improve the convergence rate of the FuLMS algorithm, it is reasonable to consider RLS-based IIR filters. The algorithms in \cite{montazeri2010a,montazeri2011a} were considered using the filtered-u RLS (FuRLS)-type algorithm for active noise and vibration control systems. Moreover, two novel fast implementation schemes were proposed for the FuRLS algorithm, generating the fast FuRLS algorithm. Simulation results demonstrated that two fast FuRLS algorithms outperform the FuLMS algorithm and SHARF algorithm.

The contributions of the existing works for broadband noise are summarized in Table \ref{BA}.
\begin{table}[h]
    \scriptsize
    \centering \doublerulesep=0.05pt
    \caption{Contributions of the algorithm for broadband noises in the past decade.}
    \begin{tabular}{l|l|l|l}
        \hline\hline
        \begin{tabular}[c]{@{}l@{}}\textbf{FxLMS-based}\\ \textbf{algorithms}\end{tabular} & \begin{tabular}[c]{@{}l@{}}\textbf{FxAP-based}\\ \textbf{algorithms}\end{tabular}  & \begin{tabular}[c]{@{}l@{}}\textbf{FxRLS-based}\\ \textbf{algorithms}\end{tabular} &
        \begin{tabular}[c]{@{}l@{}}\textbf{Subband ANC}\\ \textbf{algorithms}\end{tabular} \\ \hline
        \cite{ardekani2010theoretical,Iman2013Stochastic,chan2012performance,miyoshi2011statistical-mechanics,miyoshi2015statistical,murata2017statistical,ardekani2011stability,TabatabaeiRoot,ardekani2013stability,ardekani2015efficient,AlFiltered,cheer2016active,shi2019optimal,lu2017improved,padhi2018a,qiu2016multi,rout2015computationally,tang2012time,zhang2019a,chan2012new,zecevic2015frequency,zheng2013blind}
        &\cite{ferrer2012steady-state,ferrer2011transient,song2015an,guo2020convergence,guomean}                                                                 &\cite{reddy2011hybrid}
        &\cite{chu2015a,milani2009a,milani2010analysis,cheer2017an}                                                 \\ \hline
        \begin{tabular}[c]{@{}l@{}}\textbf{Lattice ANC}\\ \textbf{algorithms}\end{tabular} & \begin{tabular}[c]{@{}l@{}}\textbf{FeLMS-based}\\ \textbf{algorithms}\end{tabular} & \begin{tabular}[c]{@{}l@{}}\textbf{FuLMS-based}\\ \textbf{algorithms}\end{tabular} &
        \begin{tabular}[c]{@{}l@{}}\textbf{FuRLS-based}\\ \textbf{algorithms}\end{tabular}  \\ \hline
        \cite{chen2015improving,kim2012a,kim2013a,kim2020recursive}
        &\cite{lopez2009modified}                                                                 &\cite{landau2013iir,kim2011modified-filtered-u,park2012a,ho2020development}
        &\cite{montazeri2010a,montazeri2011a}                                                                \\ \hline\hline
    \end{tabular}
\label{BA}
\end{table}

\subsection{Computational complexity}
    \begin{table}[h]
        \scriptsize
        \centering \doublerulesep=0.05pt
        \caption{Computational complexity of the algorithms.}
    \begin{tabular}{l|l|l}
        \hline
        \textbf{Types}                                                                          & \textbf{Algorithms}               & \textbf{Number of multiplications} \\ \hline
        \multirow{5}{*}{\begin{tabular}[c]{@{}l@{}}Filtered-x\\ ANC family\end{tabular}} & FxLMS                             & $2M+L_s+1$                          \\ \cline{2-3}
        & FxAP                              &$2P^2M+2PM+M+L_s$                                    \\ \cline{2-3}
        & FxRLS                             &$3M^2+5M+L_s+2$                                    \\ \cline{2-3}
        & Subband ANC                 &$3M+NL_a+2(L_a+1)+L_s$                                    \\ \cline{2-3}
        & FxGAL                            & $21M+2L_s$                          \\ \hline
        \multirow{2}{*}{\begin{tabular}[c]{@{}l@{}}Filtered-e\\ ANC family\end{tabular}} & FeLMS (ALMS)                &$2M+L_s+1$                                    \\ \cline{2-3}
        &FeLMS (SPE) &$2M+L_p+1$                                    \\ \hline
        \begin{tabular}[c]{@{}l@{}}Filtered-u\\ ANC \\ family\end{tabular}                  & FuLMS                             &$2(L_f+L_b)+L_s+1$                                    \\ \hline
    \end{tabular}
\label{COM}
\end{table}

In this subsection, the computational complexity of the classical algorithms is analyzed. The number of multiplications for each algorithm per iteration is utilized to estimate the computational complexity. The computational complexity of some classical algorithms is summarized in Table \ref{COM}, where $L_s$ denotes the length of the secondary path model, $L_p$ denotes the length of the  (pseudo-)inverse of secondary path model, $L_a$ denotes the length of analysis
and synthesis filters, and $L_f$ and $L_b$ are the length of the forward and
feedback sections of the IIR filter \footnote{For computational complexity of the multi-channel ANC algorithms, the reader can refer to \cite{debrunner2006hybrid,bouchard2000}.} One can observe from this table that the LMS-based algorithms require the smallest computational complexity as compared to other algorithms and the RLS-type algorithms have the largest  computational load.

\section{Practical considerations}
\label{Pra}
\begin{table}[h]
    \scriptsize
    \centering \doublerulesep=0.05pt
    \caption{Fundamental of the online secondary path estimation before the past decade.}
    \begin{tabular}{l|l|l|l}
        \hline
        \textbf{Years} & \textbf{Authors} & \textbf{Contributions} & \textbf{References} \\ \hline
        1989       &\begin{tabular}[c]{@{}l@{}}Eriksson and\\ Allie\end{tabular}         &\begin{tabular}[c]{@{}l@{}}Online secondary path\\ estimation by injecting\\ a random noise\end{tabular}               &\begin{tabular}[c]{@{}l@{}} \cite{eriksson1989use}    \end{tabular}       \\ \hline
        1993       &\begin{tabular}[c]{@{}l@{}}Bao, Sas,\\ and Brussel\end{tabular}         &\begin{tabular}[c]{@{}l@{}}Using an additional filter\\ to reduce the interference\\ caused by injected noise \end{tabular}               &\begin{tabular}[c]{@{}l@{}} \cite{bao1993adaptive}    \end{tabular}       \\ \hline
        1997       &\begin{tabular}[c]{@{}l@{}}Kuo and Vijayan\end{tabular}         &\begin{tabular}[c]{@{}l@{}}Using an additional filter\\ to reduce the interference\\ in the secondary path estimation \end{tabular}               &\begin{tabular}[c]{@{}l@{}} \cite{kuo1997secondary}    \end{tabular}       \\ \hline
        2001       &\begin{tabular}[c]{@{}l@{}}Zhang, Lan,\\ and Ser\end{tabular}         &\begin{tabular}[c]{@{}l@{}}Cross-updated adaptive\\ filters for online secondary\\ path estimation \end{tabular}               &\begin{tabular}[c]{@{}l@{}} \cite{zhang2001cross}    \end{tabular}       \\ \hline
        2002       &\begin{tabular}[c]{@{}l@{}}Lan, Zhang\\ and Ser\end{tabular}         &\begin{tabular}[c]{@{}l@{}}Varied auxiliary noise to reduce\\the residual noise  \end{tabular}               &\begin{tabular}[c]{@{}l@{}} \cite{lan2002active}    \end{tabular}       \\ \hline
        2005       &\begin{tabular}[c]{@{}l@{}} Zhang, Lan\\ and Ser\end{tabular}         &\begin{tabular}[c]{@{}l@{}}Auxiliary noise with power\\ scheduling strategy to reduce\\the residual noise  \end{tabular}               &\begin{tabular}[c]{@{}l@{}} \cite{zhang2005comparison}    \end{tabular}       \\ \hline
        2006       &\begin{tabular}[c]{@{}l@{}}Akhtar, Abe,\\ and Kawamata\end{tabular}         &\begin{tabular}[c]{@{}l@{}}Modified-FxLMS (MFxLMS)\\ with online secondary path\\ estimation \end{tabular}               &\begin{tabular}[c]{@{}l@{}} \cite{akhtar2006new}    \end{tabular}       \\ \hline
        2008       &\begin{tabular}[c]{@{}l@{}}Carini and\\ Malatini\end{tabular}         &\begin{tabular}[c]{@{}l@{}}Optimal VSS and auxiliary noise\\ with self-tuning power scheduling \end{tabular}               &\begin{tabular}[c]{@{}l@{}} \cite{carini2008optimal}    \end{tabular}       \\ \hline
    \end{tabular}
\label{OL}
\end{table}

\subsection{Online secondary path estimation}

In the previously described works and other similar references on the topic, the solutions typically rely on the assumption $S(z)=\hat S(z)$. This assumption is based on the fact that the estimate $\hat S(z)$ can be very similar to the true value $S(z)$, depending on the estimation method used, the equipment, and the room conditions\footnote{It should be highlighted that the ANC system with heuristic algorithm usually does not require the estimation of $S(z)$, see Part II of this work.}. However, in many cases, the secondary path is time-varying. The performance may deteriorate when the secondary path varies after offline secondary path estimation. Therefore, there is a need for online secondary path estimation. The estimation of the secondary path has been addressed in previous studies, including \cite{lopes2017mmfxlms,chang2014feedforward,padhi2018performance,aslam2015new}. The fundamental of online secondary path estimation before the past decade is summarized in Table \ref{OL}. The effect of the secondary path estimation error on the performance of ANC systems was presented in \cite{ardekani2012effects}. Interestingly, it showed that the imperfect secondary path models can improve the convergence speed, but shrink the stability bound and degrade the steady-state performance. The algorithm in \cite{aslam2016maximum} comprised the least squares (LS) method and the maximum likelihood (ML) principle, which requires only one parameter. However, as shown in \cite{ardekani2015statistical}, the classical statistical estimation method, such as ML, is not stable for secondary path estimation. To overcome this limitation, a Bayesian method of maximum a posteriori was suggested, which gives a feasible and stable solution to the secondary path estimation problem \cite{ardekani2015statistical}. In \cite{gaiotto2013a}, a tuning-less approach was derived from the method of least squares (LS). The adaptation of this rule is completely devoid of parameters, leading to easy implementation.

 In addition to decreasing the number of parameters, the update ratio also plays an important role in reducing complexity. A novel filtered-x LMS-Newton algorithm was proposed, by extending the LMS-Newton algorithm to ANC systems \cite{aslam2018variable}. The LMS-Newton algorithm can converge faster than the conventional LMS algorithm when the input signal is highly correlated, and it is mathematically identical to the RLS algorithm when setting $2\mu=1-\lambda$ \cite{poulo2008adaptive}. Moreover, the selective updating scheme was incorporated in the filtered-x LMS-Newton algorithm to reduce computational complexity of the adaptation of controller. Similar to the set-membership filtering, the selective updating scheme involves a predefined threshold to determine the feasibility solution set. For these reasons, a variable threshold strategy was further developed for performance improvement \cite{aslam2018variable}.

 The ANC system with online secondary path estimation via auxiliary noise injection method has been extensively studied. In this method, the white Gaussian noise with zero mean is commonly used as the auxiliary noise signal. Such auxiliary noise signal is employed as the input to a classical system identification problem. However, this injected auxiliary noise can deteriorate the noise reduction performance. To overcome this problem, the variable power methods of auxiliary noise were preferred \cite{ahmed2013robust,lopes2015auxiliary}. In \cite{lopes2015auxiliary}, an auxiliary noise-based method was designed to tackle sudden and strong changes of $S(z)$, where a mixture of the LMS and normalized LMS (NLMS) algorithms is adopted for secondary path estimation. Alternatively, the VSS scheme can be employed for combating performance degradation and sudden changes of $S(z)$ \cite{HaseebAp,kim2020two}. In \cite{kim2020two}, a scheduled-step size NLMS algorithm was employed to estimate the secondary path while suppressing disturbances from the modeling and controlling filter. In \cite{yang2018online}, both variable power and VSS strategies were employed for performance improvement.

 In NANC systems, the problem of online estimation of the secondary path also needs to be considered \cite{delega2017a,ma2016new}. It was demonstrated that such mismatch between $S(z)$ and $\hat S(z)$ can affect the stable range of step size \cite{wang2009convergence}. In \cite{chang2018secondary}, the effect of both offline and online secondary path modeling was examined by theoretical analysis, which shows that such imperfection can slow down the convergence rate. Then, it offered a solution to speed up the convergence. To provide an online secondary path estimation method for NANC systems, a modified FxLMS algorithm with secondary path estimation was proposed, which is based on scaled auxiliary noise injection \cite{liu2010analysis}. A novel mirror-FxLMS algorithm was proposed in \cite{Paulo2019A}, which is based on the algorithm in \cite{lopes2017mmfxlms} and can guarantee stability without an auxiliary signal to online estimate secondary path. Such an algorithm can switch between the modified FxLMS (MoFxLMS) and mirror-modified FxLMS (MMoFxLMS) algorithms according to the absolute value of their weight coefficients\footnote{Here, the modified FxLMS is referred to the algorithm proposed by Rupp and Sayed \cite{rupp1995modified}. To avoid confusion, we abbreviate this algorithm as MoFxLMS, and the modified FxLMS algorithm proposed by Akhtar is termed as MFxLMS algorithm \cite{akhtar2006new}.}. If the absolute value of the MMoFxLMS algorithm greater than the MoFxLMS, the MMoFxLMS algorithm adapts the controller; otherwise, the MoFxLMS algorithm works. The MoFxLMS algorithm can obtain improved performance under high secondary path modeling error, while the MMoFxLMS algorithm can converge fast regardless of the secondary path modeling errors. As a result, a stable performance is achieved for the overall filter. In particular, a phase-locked loop (PLL) was exploited to enhance the tracking capability.

 Similarly, Akhtar combined the new algorithms with his MFxLMS algorithm for acquiring refined performance in the presence of $\alpha$-stable noise \cite{akhtar2016binormalized,akhtar2019a}. The above examples clearly demonstrate the large potential of online secondary path estimation, not only for Gaussian noise environment, but also for impulsive noise.

\subsection{Solutions of acoustic feedback}

In practical situations, the antisound output $y(n)$ to the loudspeaker also propagates upstream to the reference microphone, leading to a corrupted reference signal. The coupling of the acoustic wave from the active loudspeaker (cancelling loudspeaker) to the reference microphone is called \textit{acoustic feedback}. The development of the acoustic feedback before the past decade is outlined in Table \ref{AFB}.

\begin{table}[h]
    \scriptsize
    \centering \doublerulesep=0.05pt
    \caption{Development of the acoustic feedback before the past decade.}
        \begin{tabular}{l|l|l|l}
            \hline
            \textbf{Years} & \textbf{Authors} & \textbf{Contributions} & \textbf{References} \\ \hline
            1994       &\begin{tabular}[c]{@{}l@{}}Kuo and\\ Luan\end{tabular}         &\begin{tabular}[c]{@{}l@{}}Online feedback compensation \\ for multi-channel ANC\end{tabular}               &\begin{tabular}[c]{@{}l@{}} \cite{kuo1994line}    \end{tabular}       \\ \hline
            2002       &\begin{tabular}[c]{@{}l@{}}Kuo\end{tabular}         &\begin{tabular}[c]{@{}l@{}} Online feedback path\\ neutralization filter for ANC\end{tabular}               &\begin{tabular}[c]{@{}l@{}}\cite{kuo2002active}    \end{tabular}       \\ \hline
            2002       &\begin{tabular}[c]{@{}l@{}}Sun and,\\ Chen\end{tabular}         &\begin{tabular}[c]{@{}l@{}} Online feedback path\\ neutralization filter using new\\ FuLMS algorithm\end{tabular}               &\begin{tabular}[c]{@{}l@{}}\cite{sun2002new}    \end{tabular}       \\ \hline
            2007       &\begin{tabular}[c]{@{}l@{}}Akhtar, Abe,\\ and Kawamata\end{tabular}         &\begin{tabular}[c]{@{}l@{}}Online acoustic feedback path\\ modeling for both narrowband\\ and broadband ANC\end{tabular}               &\begin{tabular}[c]{@{}l@{}}\cite{akhtar2007active}    \end{tabular}       \\ \hline
    \end{tabular}
    \label{AFB}
\end{table}
\begin{figure}[!htb]
    \centering
    \includegraphics[scale=0.5] {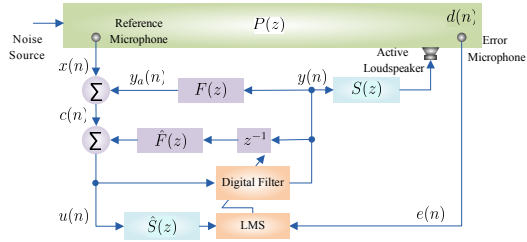}
    \caption{Diagram of the FxLMS algorithm with feedback neutralization.}
    \label{FigAFB}
\end{figure}

The \textit{feedback neutralization} is the simplest acoustic feedback solution, which uses a separate feedback path neutralization (FBPN) filter with the controller. The principle of feedback neutralization is exactly the same as that of acoustic echo cancellation \cite{benesty2001advances}. A block diagram of the ANC system using the FxLMS algorithm
with feedback neutralization is shown in Fig. \ref{FigAFB}, where $F(z)$ is the
feedback path from cancelling loudspeaker to the reference microphone, $\hat F(z)$ denotes the estimate of $F(z)$ (FBPN filter), and $y_a(n)$ denotes the acoustic feedback component. As can be seen, this electrical model of the feedback path is driven by the antinoise signal $y(n)$ , and its output is subtracted from the reference sensor signal $c(n)$. Finally, a (somewhat) acoustic-feedback-free reference
signal $u(n)$ is obtained for adaptation of the control filter. The FBPN filter can be adapted offline or online. In many applications, $F(z)$ may be time-varying. Hence, online acoustic feedback modeling and neutralization are needed.

During the past decade, several algorithms have been proposed for online acoustic feedback path modeling and neutralization \cite{tufail2019two,akhtar2009simplified,akhtar2011variable,Ahmed2013Online,ShakeelGain}. Similar to the online secondary path estimation, an auxiliary noise modeled by white Gaussian noise is injected to an ANC system for online modeling feedback path. In \cite{akhtar2011variable}, a novel VSS scheme was proposed for FBPN filter and then it extended to multi-channel ANC systems. With slightly increased computational complexity, it achieves faster convergence than that of previous algorithms. Following a different direction, the algorithms in \cite{Ahmed2013Online,ShakeelGain} developed a time-varying gain scheme for white Gaussian noise generation. Since the  auxiliary noise contributes to the residual error, the attenuated performance is degraded. By using the time-varying gain scheme, the lower noise level can be achieved.

\subsection{Measurement of error signal}

\begin{table}[h]
    \scriptsize
    \centering \doublerulesep=0.05pt
    \caption{Fundamental of the error signal measurement before the past decade.}
        \begin{tabular}{l|l|l|l}
            \hline
            \textbf{Years} & \textbf{Authors} & \textbf{Contributions} & \textbf{References} \\ \hline
            1992       &\begin{tabular}[c]{@{}l@{}}Elliott,\\ and David\end{tabular}         &\begin{tabular}[c]{@{}l@{}}Virtual microphone arrangement\end{tabular}               &\begin{tabular}[c]{@{}l@{}} \cite{elliott1992virtul}    \end{tabular}       \\ \hline
            1999       &\begin{tabular}[c]{@{}l@{}}Roure and\\ Albarrazin\end{tabular}         &\begin{tabular}[c]{@{}l@{}}Remote microphone technique \end{tabular}               &\begin{tabular}[c]{@{}l@{}}\cite{roure1999remote}    \end{tabular}       \\ \hline
            2002       &\begin{tabular}[c]{@{}l@{}}Cazzolato\end{tabular}         &\begin{tabular}[c]{@{}l@{}}LMS filter with virtual microphone technique \end{tabular}               &\begin{tabular}[c]{@{}l@{}} \cite{cazzolato2002adaptive}    \end{tabular}       \\ \hline
            2006       &\begin{tabular}[c]{@{}l@{}}Diaz, Ega{\~n}a,\\ and Vinolas\end{tabular}         &\begin{tabular}[c]{@{}l@{}}FxLMS with virtual microphone\\ for railway sleeping vehicle applications\end{tabular}               &\begin{tabular}[c]{@{}l@{}} \cite{diaz2006local}    \end{tabular}       \\ \hline
            2007       &\begin{tabular}[c]{@{}l@{}}Liao and Lin\end{tabular}         &\begin{tabular}[c]{@{}l@{}} Investigate several FIR algorithms with\\ communication error for ANC systems\end{tabular}               &\begin{tabular}[c]{@{}l@{}} \cite{liao2007new}    \end{tabular}       \\ \hline
            2008       &\begin{tabular}[c]{@{}l@{}}Petersen\\ et al.\end{tabular}         &\begin{tabular}[c]{@{}l@{}}Kalman filter with virtual sensing technique\end{tabular}               &\begin{tabular}[c]{@{}l@{}} \cite{petersen2008kalman}    \end{tabular}       \\ \hline
    \end{tabular}
    \label{EL}
\end{table}

We briefly revisit the measurement method of the error signal in this subsection. In this context, the literature is rather scarce. Some contributions of the error signal measurement before the past decade are listed in Table \ref{EL}. In \cite{wu2015a}, the concept of the \textit{communication error} $e_c(n)$ was applied to ANC algorithms, which can be expressed as
\begin{equation}
e_c(n) = e(n) - s(n)*[\bm w^{\mathrm T}(n)\bm x(n)] + \bm w^{\mathrm T}(n)\bm X(n).
\label{008a}
\end{equation}
Such measurement can be interpreted as a potential difference related with the sequence A ($\bm x(n)$ $\to$ secondary path $S(z)$ $\to$ filter $\bm w(z)$ $\to$ error signal $e(n)$) and the sequence B ($\bm x(n)$ $\to$ filter $\bm w(z)$ $\to$ secondary path $S(z)$ $\to$ error signal $e(n)$). The price paid for the better performance of the ANC algorithm with $e_c(n)$ is an increase in computational load as compared to the other measurement schemes under evaluation.

On the other hand, retaining the residual noise with specified spectrum has become necessary to offer a better natural feeling \cite{jiang2018review}. The adaptive noise equalizer (ANE) algorithm meets the requirement of attenuating or amplifying a predetermined sinusoidal noise, whose error signal is defined by
\begin{equation}
e(n) \triangleq d(n) - (1-\theta) y(n)
\label{009a}
\end{equation}
where $\theta$ denotes the gain value. For $\theta = 0$, it can attenuate the noise source completely; for $\theta = 0.5$, it can reduce the amplitude of the noise source by half; for $\theta = 1$, the amplitude of noise source is unchanged; for $\theta = 2$, the ANE amplifies the amplitude of the noise source by 2. By doing so, such method can meet all environments' requirements in a more flexible manner.

\begin{figure}[!htb]
    \centering
    \includegraphics[scale=0.5] {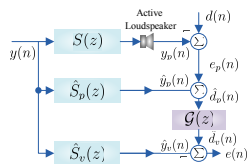}
    \caption{\label{08} Estimation model of conventional virtual ANC system.}
    \label{Fig08}
\end{figure}

Alternatively, a new measurement of the error signal can be derived by implementing the virtual sensing systems \cite{vicente2013effect,das2013computationally}. Among them, the remote microphone technique (RMT) is the most effective virtual sensing algorithm, which employs the offline identification of two secondary paths. One estimated transfer function of physical location $\hat S_p(z)$ is defined according to the path between the control signal and the physical microphone position, while the other estimated transfer function of $\hat S_v(z)$ represents the virtual locations between the control signal and the virtual microphone position. The diagram of RMT can be seen in Fig. \ref{Fig08}, where $e(n)$ can be interpreted as the total error signal at the virtual location, $e_p(n)$ denotes the error signal from a physical microphone, $\hat y_p(n)$ and $\hat y_v(n)$ is the estimation of the secondary disturbance at the physical and virtual microphones, respectively, and $\mathcal G(z)$ denotes the transfer function between the physical and virtual locations. The error signal in this model can be defined by
\begin{equation}
e(n) \triangleq d_v(n) - \hat s_v(n)*\hat y_v(n)
\label{009aa}
\end{equation}
where $\hat s_v(n)$ is the impulse response of $\hat S_v(z)$ and $d_v(n)$ denotes the primary disturbance at the virtual location. Due to the additional calculations from three transfer functions, $\hat S_p(z)$, $\hat S_v(z)$, and $\mathcal G(z)$, the FxLMS algorithm with virtual  microphone (virtual FxLMS algorithm) has increased computational complexity as compared to the FxLMS algorithm. To reduce computational complexity,  the algorithm in \cite{das2013computationally} provides a solution by resorting to the frequency-domain method, which also has better performance than the traditional virtual FxLMS algorithm.

To evaluate the performance of the ANC algorithm, another idea is to utilize the acoustic pressure (sound pressure), which is estimated based on the acceleration measurements at the stages of tuning and operating \cite{gonzalez2003sound}.

\subsection{Frequency mismatch}

In 2006, Xiao et al. proposed that the actual frequency of the primary noise and the frequency of the synthesized reference signal in NANC systems may be mismatch \cite{xiao2006new}. To explore the effect of mismatch, the FxLMS algorithm was analyzed from the perspective of frequency mismatch and phase error \cite{jeon2009analysis}. The results showed that the ANC system may undergo severe performance degradation even if the frequency mismatch is only 1\%. Several algorithms were proposed to compensate for this mismatch, including \cite{jeon2010narrowband,liu2017analysis,han2019narrowband}.
The ANF can provide accurate estimation of frequency and it has been used to enhance performance \cite{wang2014new}. In \cite{wang2019narrowband}, a parallel ANF was proposed, where multiple tones were separated in the feedback reconstruction process to reduce the negative effect of frequency interference.

\subsection{Design of feedback ANC systems}

\begin{table}[h]
    \scriptsize
    \centering \doublerulesep=0.05pt
    \caption{Contributions of feedback ANC systems before the past decade.}
        \begin{tabular}{l|l|l|l}
            \hline
            \textbf{Years} & \textbf{Authors} & \textbf{Contributions} & \textbf{References} \\ \hline
            1991       &\begin{tabular}[c]{@{}l@{}}Eriksson\end{tabular}         &\begin{tabular}[c]{@{}l@{}}Feedback ANC system \end{tabular}               &\begin{tabular}[c]{@{}l@{}}\cite{lj1991ii}    \end{tabular}       \\ \hline
            1992       &\begin{tabular}[c]{@{}l@{}}Popovich, Melton,\\ and Allie\end{tabular}         &\begin{tabular}[c]{@{}l@{}}Multi-channel feedback\\ ANC system \end{tabular}               &\begin{tabular}[c]{@{}l@{}}\cite{popovich1992new}    \end{tabular}       \\ \hline
            1997       &\begin{tabular}[c]{@{}l@{}}Bai and Lee\end{tabular}         &\begin{tabular}[c]{@{}l@{}}Using $H_\infty$ optimization technique to\\ design the feedback ANC system\end{tabular}               &\begin{tabular}[c]{@{}l@{}} \cite{bai1997implementation}    \end{tabular}       \\ \hline
            1999       &\begin{tabular}[c]{@{}l@{}}Rafaely and\\ Elliott\end{tabular}         &\begin{tabular}[c]{@{}l@{}}Using $H_2$/$H_\infty$ method to design\\ the feedback ANC system and apply\\ to headrest systems\end{tabular}               &\begin{tabular}[c]{@{}l@{}} \cite{rafaely1999h}    \end{tabular}       \\ \hline
            1999       &\begin{tabular}[c]{@{}l@{}}Chen, Chiueh,\\ and Chen\end{tabular}         &\begin{tabular}[c]{@{}l@{}}Feedback ANC for magnetic\\ resonance noise control\end{tabular}               &\begin{tabular}[c]{@{}l@{}} \cite{chen1999active}    \end{tabular}       \\ \hline
            2003       &\begin{tabular}[c]{@{}l@{}}Kuo, Kong,\\ and Gan\end{tabular}         &\begin{tabular}[c]{@{}l@{}}Feedback ANC algorithm with three\\ distributed error sensors for industrial\\ machine noise control\end{tabular}               &\begin{tabular}[c]{@{}l@{}}\cite{kuo2003applications}    \end{tabular}       \\ \hline
            2008       &\begin{tabular}[c]{@{}l@{}}Zhou et al.\end{tabular}         &\begin{tabular}[c]{@{}l@{}}Feedback ANC system using a model-free\\ controller with simultaneous\\ perturbation stochastic approximation \end{tabular}               &\begin{tabular}[c]{@{}l@{}}\cite{zhou2008use}    \end{tabular}       \\ \hline
    \end{tabular}
    \label{FB}
\end{table}

As we discussed in the Introduction, the feedback ANC systems do not require a coherent reference signal and can be used in the case that the reference signal is not available. The main contributions of the feedback ANC systems before the past decade are summarized in Table \ref{FB}. During the past decade, such systems have been widely used to ANC headphones and fMRI acoustic noise control, etc (see `Recent implementations and applications of ANC' in Part II).

The feedback ANC systems can be divided into two categories: non-adaptive systems and adaptive systems. The non-adaptive systems can effectively attenuate noise with frequency band of interest by a fixed controller. Unfortunately, the non-stationary reference signals bring great challenges to the stability of such systems. In contrast, the feedback ANC systems with adaptive controller can deal with non-stationary reference signals and the stability problems are no longer a great challenge for adaptive systems. In \cite{wu2018a}, a generalized form of the LFxLMS algorithm was proposed for feedback ANC systems, which replaces $\gamma$ with a designed symmetric Toeplitz matrix $\bm Q$. In \cite{wu2014a}, a simplified feedback ANC system was proposed which employs the residual noise directly as the reference signal. Compared to the IMC algorithm \cite{Tongwei2014Stochastic}, it shows a lower computational burden since it does not compute the convolution.

\subsection{Analog control in ANC}

    \begin{table}[h]
        \scriptsize
        \centering \doublerulesep=0.05pt
        \caption{Development of the analog control before the past decade.}
            \begin{tabular}{l|l|l|l}
                \hline
                \textbf{Years} & \textbf{Authors} & \textbf{Contributions} & \textbf{References} \\ \hline
                1956       &\begin{tabular}[c]{@{}l@{}}Hawley \end{tabular}         &\begin{tabular}[c]{@{}l@{}}ANC headset with an analog controller \end{tabular}               &\begin{tabular}[c]{@{}l@{}} \cite{hawley1956acoustic}    \end{tabular}       \\ \hline
                1987       &\begin{tabular}[c]{@{}l@{}}Leitch and\\ Tokhi \end{tabular}         &\begin{tabular}[c]{@{}l@{}}A review for ANC systems, including analog control\\ methods \end{tabular}               &\begin{tabular}[c]{@{}l@{}} \cite{leitch1987active}    \end{tabular}       \\ \hline
                2001       &\begin{tabular}[c]{@{}l@{}}Yu and Hu \end{tabular}         &\begin{tabular}[c]{@{}l@{}}An active analog filter is used to realize the fourth-order\\ controller \end{tabular}               &\begin{tabular}[c]{@{}l@{}} \cite{yu2001controller}    \end{tabular}       \\ \hline
                2002       &\begin{tabular}[c]{@{}l@{}}Pawelczyk \end{tabular}         &\begin{tabular}[c]{@{}l@{}}A detailed procedure for designing and practically\\ realizing an analogue ANC system \end{tabular}               &\begin{tabular}[c]{@{}l@{}} \cite{pawelczyk2002analogue}    \end{tabular}       \\ \hline
                2005       &\begin{tabular}[c]{@{}l@{}}Song, Gong,\\ and Kuo \end{tabular}         &\begin{tabular}[c]{@{}l@{}}Adding an analog feedback loop to a digital ANC system,\\ generating a hybrid feedback ANC headset \end{tabular}               &\begin{tabular}[c]{@{}l@{}} \cite{song2005robust}    \end{tabular}       \\ \hline
        \end{tabular}
        \label{ANA}
    \end{table}

In the feedback ANC systems, analog control (analog circuitry) based on a negative feedback loop is broadly used for headphones, due to the cost and battery-life issues \cite{bai1997implementation}. It is known that utilizing both digital and analog control has the ability to cancel broadband and narrowband noises. In particular, the analog control is fairly inexpensive and it has good broadband noise reduction owing to the short time delay of analog components employed in the controller \cite{song2005robust}. However, the performance of the analog controller for NANC is limited because it is unable to track the environmental changes, which may hinder the application of analog control in some cases \cite{song2005robust}. To address this limitation, a hybrid system was proposed in \cite{song2005robust}, which adds an analog feedback loop into a digital ANC system to reduce disturbances. We summarize the development of analog control for ANC in Table \ref{ANA}.

Note that some important analog control methods have been proposed before the past decade, and relatively few studies have been conducted in these years. In \cite{benois2018hybrid}, a short review was presented for hybrid analog-digital systems, including six possible combinations
of the classical control structures for ANC systems. The development of the analog control in recent years can also be found in \cite{schumacher2011active,nagahara2013active}. In \cite{nagahara2013active}, a continuous FxLMS algorithm was proposed for hybrid analog-to-digital systems. Moreover, an approximation of the algorithm was developed, which can be easily implemented in digital signal processor (DSP).

\subsection{Methods of reducing computational complexity}

\begin{table}[h]
    \scriptsize
    \centering \doublerulesep=0.05pt
    \caption{Fundamental of the reduced computational complexity for ANC before the past decade.}
        \begin{tabular}{l|l|l|l}
            \hline
            \textbf{Years} & \textbf{Authors} & \textbf{Contributions} & \textbf{References} \\ \hline
            1995       &\begin{tabular}[c]{@{}l@{}}Douglas\end{tabular}         &\begin{tabular}[c]{@{}l@{}}Fast FxAP algorithm \end{tabular}               &\begin{tabular}[c]{@{}l@{}} \cite{douglas1995fast}    \end{tabular}       \\ \hline
            1999       &\begin{tabular}[c]{@{}l@{}} Douglas \end{tabular}         &\begin{tabular}[c]{@{}l@{}}Fast multi-channel FxLMS algorithm\end{tabular}             &\begin{tabular}[c]{@{}l@{}} \cite{douglas1999fast}\end{tabular}            \\ \hline
            2003       &\begin{tabular}[c]{@{}l@{}} Bouchard \end{tabular}         &\begin{tabular}[c]{@{}l@{}}Fast multi-channel FxAP algorithm for ANC\\ and  and acoustic equalization systems\end{tabular}             &\begin{tabular}[c]{@{}l@{}} \cite{bouchard2003multichannel}\end{tabular}            \\ \hline
            2006       &\begin{tabular}[c]{@{}l@{}}Carini and\\ Sicuranza\end{tabular}         &\begin{tabular}[c]{@{}l@{}}Multi-channel FxAP algorithm\\ with set-membership filtering \end{tabular}               &\begin{tabular}[c]{@{}l@{}}\cite{carini2006analysis}    \end{tabular}       \\ \hline
            2007       &\begin{tabular}[c]{@{}l@{}}Albu, Bouchard,\\ and Zakharov\end{tabular}         &\begin{tabular}[c]{@{}l@{}}Filtered-x pseudo AP algorithm\\ based on Gauss-Seidel method\\ or dichotomous coordinate descent (DCD) \end{tabular}               &\begin{tabular}[c]{@{}l@{}}\cite{albu2007pseudo}    \end{tabular}       \\ \hline
            2008       &\begin{tabular}[c]{@{}l@{}}Wesselink and\\ Berkhoff\end{tabular}         &\begin{tabular}[c]{@{}l@{}}Filtered-e fast algorithms for multi-channel ANC \end{tabular}               &\begin{tabular}[c]{@{}l@{}} \cite{wesselink2008fast}    \end{tabular}       \\ \hline
    \end{tabular}
    \label{ST}
\end{table}

As an important structure of ANC systems, the subband ANC algorithm is well-suited for multi-channel, MIMO or large-scale ANC systems while preserving the conceptual simplicity of the classical SAFs. The graphics processing units (GPUs) offer powerful parallel processing and it also has reduced computational demand in the previous studies \cite{lorente2015the,lorente2014gpu}. Some algorithms for reducing complexity can also be integrated into the classic ANC algorithm, such as fast algorithms, \textit{set-membership filtering} and PU, moreover, the PU scheme has been applied to modify the FxLMS and FxAP algorithm \cite{xiao2016a,lu2012fixed}. Note that the first attempt to simultaneously apply PU and set-membership approaches in ANC algorithms is presented for a NLANC system, but not for a linear system \cite{sicuranza2005nonlinear}. We reviewed these works in Part II of this work, see Table 4 in \cite{lu2020survey}. Before the past decade, several linear ANC algorithms were developed by using these schemes, see Table \ref{ST}. In future research, applying both algorithms and structures to reduce complexity can be considered by resorting to research results presented in the literature before the last decade.

\subsection{Active structural acoustic control (ASAC)}

\begin{table}[h]
    \scriptsize
    \centering \doublerulesep=0.05pt
    \caption{Time line of the ASAC before the past decade.}
    \begin{tabular}{l|l|l|l}
        \hline
        \textbf{Years} & \textbf{Authors} & \textbf{Contributions} & \textbf{References} \\ \hline
        1987       &\begin{tabular}[c]{@{}l@{}}Fuller and Jones \end{tabular}         &\begin{tabular}[c]{@{}l@{}}Using active vibration control of aircraft\\ fuselages to reduce interior noise levels \end{tabular}               &\begin{tabular}[c]{@{}l@{}} \cite{fuller1987experiments}    \end{tabular}       \\ \hline
        1991       &\begin{tabular}[c]{@{}l@{}}Sommerfeldt \end{tabular}         &\begin{tabular}[c]{@{}l@{}}Multi-channel FxLMS algorithm was applied\\ to vibrational ANC \end{tabular}               &\begin{tabular}[c]{@{}l@{}} \cite{sommerfeldt1991multi}    \end{tabular}       \\ \hline
        1992       &\begin{tabular}[c]{@{}l@{}}Clark and Fuller \end{tabular}         &\begin{tabular}[c]{@{}l@{}}Investigate the implementation of the error sensor\\ constructed from polyvinylidene fluoride (PVDF)\\ for ASAC \end{tabular}               &\begin{tabular}[c]{@{}l@{}} \cite{clark1992optimal,clark1992modal}    \end{tabular}       \\ \hline
        1992       &\begin{tabular}[c]{@{}l@{}}Baumann, Ho, \\and Robertshaw \end{tabular}         &\begin{tabular}[c]{@{}l@{}}ASAC for broadband disturbances \end{tabular}               &\begin{tabular}[c]{@{}l@{}} \cite{baumann1992active}    \end{tabular}       \\ \hline
        1994       &\begin{tabular}[c]{@{}l@{}}Fuller and \\Gibbs\end{tabular}         &\begin{tabular}[c]{@{}l@{}}Using small patch type piezoceramic actuators\\ bonded to fuselages for interior noise control \end{tabular}               &\begin{tabular}[c]{@{}l@{}} \cite{fuller1994active}    \end{tabular}       \\ \hline
        2000       &\begin{tabular}[c]{@{}l@{}}Gibbs et al.\end{tabular}         &\begin{tabular}[c]{@{}l@{}}MIMO adaptive sensor for ASAC \end{tabular}               &\begin{tabular}[c]{@{}l@{}} \cite{gibbs2000radiation}    \end{tabular}       \\ \hline
        2000       &\begin{tabular}[c]{@{}l@{}}Berkhoff\end{tabular}         &\begin{tabular}[c]{@{}l@{}}Sensor scheme design for  ASAC \end{tabular}               &\begin{tabular}[c]{@{}l@{}} \cite{berkhoff2000sensor}    \end{tabular}       \\ \hline
        2001       &\begin{tabular}[c]{@{}l@{}}Gardonio et al.\end{tabular}         &\begin{tabular}[c]{@{}l@{}}A theoretical and experimental study of the\\ frequency response function of a matched volume\\ velocity sensor and uniform force actuator for ASAC \end{tabular}               &\begin{tabular}[c]{@{}l@{}} \cite{gardonio2001analysis}    \end{tabular}       \\ \hline
        2004       &\begin{tabular}[c]{@{}l@{}}Carneal and\\ Fuller\end{tabular}         &\begin{tabular}[c]{@{}l@{}}An ASAC approach for double panel systems \end{tabular}               &\begin{tabular}[c]{@{}l@{}} \cite{carneal2004analytical}    \end{tabular}       \\ \hline
    \end{tabular}
    \label{VB}
\end{table}

To reduce device and machinery noises, an important approach is to control vibration of their casings \cite{misol2012experimental}. This approach is also useful from the perspective of global noise control. A number of theoretical and experimental works have been developed for ASAC. The multi-channel ANC algorithms have been widely used to control structural vibration \cite{sommerfeldt1991multi,zhao2015experimental}. Before the past decade, many methods have been proposed for ASAC. We outline some important contributions in Table \ref{VB}.

In the past decade, the ASAC techniques have been extensively used in diverse fields (helicopters \cite{belanger2009multi,ma2017active}, vehicles \cite{guo2019vehicle}, etc \cite{pinte2010piezo,ma2014active,hendricks2014experimental}). In \cite{pinte2009active}, an ASAC approach was proposed for repetitive impact noises. The efficiency of the suggested optimal control configuration and the iterative learning control (ILC) algorithm was verified. In \cite{mazur2018design}, an implementation of multi-channel global ANC systems was presented, which utilizes an active casing. Moreover, a distributed version of the switched-error FxLMS algorithm was developed as the control algorithm.

\subsection{Fuzzy control}
\begin{table}[h]
    \scriptsize
    \centering \doublerulesep=0.05pt
    \caption{Development of the fuzzy control for ANC before the past decade.}
    \begin{tabular}{l|l|l|l}
        \hline
        \textbf{Years} & \textbf{Authors} & \textbf{Contributions} & \textbf{References} \\ \hline
        1993, 1994       &\begin{tabular}[c]{@{}l@{}}Kipersztok and\\ Hammond\end{tabular}         &\begin{tabular}[c]{@{}l@{}}Fuzzy-logic system for  broadband\\ noise control \end{tabular}               &\begin{tabular}[c]{@{}l@{}} \cite{kipersztok1993active,kipersztok1994fuzzy}    \end{tabular}       \\ \hline
        1995       &\begin{tabular}[c]{@{}l@{}}Kipersztok and\\ Hammond\end{tabular}         &\begin{tabular}[c]{@{}l@{}} Improved fuzzy-logic system for\\ ANC \end{tabular}               &\begin{tabular}[c]{@{}l@{}} \cite{kipersztok1995use}    \end{tabular} \\  \hline
        2000       &\begin{tabular}[c]{@{}l@{}}Silva et al. \end{tabular}         &\begin{tabular}[c]{@{}l@{}} Fuzzy modeling techniques for\\ weak nonlinearities \end{tabular}               &\begin{tabular}[c]{@{}l@{}} \cite{silva2000acoustic}    \end{tabular}    \\ \hline
        2001, 2003       &\begin{tabular}[c]{@{}l@{}}Sousa et al. \end{tabular}         &\begin{tabular}[c]{@{}l@{}} Using direct and inverse TS fuzzy models for\\ (nonlinear) ANC \end{tabular}               &\begin{tabular}[c]{@{}l@{}} \cite{silva2001inverse,sousa2003fuzzy}    \end{tabular}    \\ \hline
        2005       &\begin{tabular}[c]{@{}l@{}}Botto, Sousa, and \\S{\'a} da Costa \end{tabular}         &\begin{tabular}[c]{@{}l@{}} Fuzzy and neural modeling techniques \\for (nonlinear) ANC \end{tabular}               &\begin{tabular}[c]{@{}l@{}} \cite{botto2005intelligent}    \end{tabular}    \\ \hline
    \end{tabular}
    \label{FFU}
\end{table}

The fuzzy control is one of the most appealing methods where it is impossible to sufficiently well model the inference system and the FxLMS algorithm might be failed. The early work of fuzzy control in ANC systems can be found in \cite{kipersztok1993active,kipersztok1994fuzzy}, which employ the fuzzy-logic system for broadband noise control. In \cite{kipersztok1995use}, the cross correlation between reference and error microphones signals and the signal-to-noise ratio (SNR) estimate of the cross-correlation function were used as the input of a fuzzy-logic system to adjust the coefficients of the FIR filter.

Before the past decade, the inverse fuzzy modeling techniques were investigated for ANC systems \cite{silva2001inverse,sousa2003fuzzy,botto2005intelligent}. In particular, such techniques allow for further performance improvements  in linear time invariant models. These methods generally use the \textit{Takagi-Sugeno} (TS) model to deal with reverberations and nonlinear distortions in ANC systems. The development of fuzzy control before the past decade is summarized in Table \ref{FFU}. In 2019, the fuzzy control methods were applied to the vehicle noise and vibration control \cite{guo2019vehicle}. It demonstrated that such method has the ability to tolerate the uncertainties of the input vibration signals and to handle the nonlinear phenomena.

It should be noted that some of the above mentioned methods are applied to NLANC systems. Although classical FIR filters achieve stable performance in these systems, the fuzzy modeling methods can perform identification more accurately. In addition, fuzzy-logic systems can also be combined with artificial neural networks (ANNs), generating fuzzy ANNs. We review the fuzzy ANNs methods in Part II.

\section{Novel linear ANC methods emerging in the past decade}
\label{sec:4}

\subsection{Psychoacoustic ANC systems}

The human hearing sensation has selective sensitivity to different frequencies. Hence, it is reasonable to take the characteristics of human hearing into account. In other words, more considerations should be given to the frequency response of the controller. Moreover, minimizing the perceived annoyance of human hearing with residual noise, which needs to be resolved.

To tackle this problem, the psychoacoustic ANC (PANC) systems were established by weighting of the reference and the error signal \cite{bao2010psychoacoustic,wang2012psychoacoustic,munir2020fxlms}. These PANC systems have similar structure to the FxFeLMS algorithm in Fig. \ref{Fig06} but with different design of the error filter $H(z)$. Besides, instead of sound pressure level (SPL) and averaged noise reduction (ANR), PANC is inclined to utilize loudness as the measurement of the performance. In \cite{wang2012psychoacoustic}, a hybrid PANC system was proposed, which can simultaneously control either uncorrelated disturbance or correlated primary noise. In \cite{belyi2019integrated}, a novel approach integrating subband PANC and psychoacoustic masking was proposed, resulting in reduced computational cost and improved perceptual sound quality and high-frequency noise reduction level.

\subsection{Sparse ANC algorithms}

Taking advantage of the sparsity that may exist in ANC systems to improve performance is a method worthy of attention. A filtered-x improved proportionate NLMS (FxIPNLMS) algorithm was devised by extending the proportionate algorithm to feedforward ANC system \cite{arenas2011combinations}. Moreover, the FxIPNLMS algorithm has been shown to be compatible with convex combination schemes for Gaussian noise source, yielding enhanced performance under different degrees of sparsity \cite{arenas2011combinations}. By using the framework of the FxAP algorithm, some algorithms were proposed which incorporates the zero-attracting (ZA) or reweighted zero-attracting (RZA) strategies \cite{albu2017low-complexity,gully2014sparsity,albu2014sparsity}. Unlike the above mentioned algorithms for the case where the primary or secondary path is sparse, the work in \cite{zhangg2016sparse} investigates the scenario where the noise source is sparsely distributed. The main concept of the proposed complex algorithm is still based on the ZA scheme and the FxLMS algorithm. The significant noise reduction and fast convergence of the algorithm were obtained via experimental tests\footnote{The involved spatial ANC technique can be seen in Part II of this work.}.

\subsection{Convex combination ANC algorithms}
\begin{figure}[!htb]
    \centering
    \includegraphics[scale=0.5] {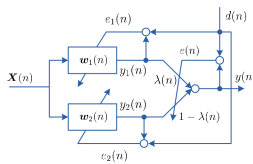}
    \caption{\label{11} Diagram of the convex combination.}
    \label{Fig11}
\end{figure}

Like the VSS schemes, the convex combination is a scheme to address the trade-off of the fast convergence rate and small noise residue caused by fixed step size, which utilizes the filter bank to simultaneously obtain refined performance \cite{ferrer2013convex,padhi2018a,arenas2011combinations}. The schematic of the convex combination is shown in Fig. \ref{Fig11}, where $e_1(n)$ denotes the error signal of the fast filter (the filter with large step size), $e_2(n)$ denotes the error signal of the slow filter (the filter with small step size), $y_1(n)$ is the output of the fast filter, $y_2(n)$ is the output of the slow filter, $\bm w_1(n)$ is the weight vector of the fast filter, $\bm w_2(n)$ is the weight vector of the slow filter, and $\lambda(n) \in [0,1]$ is the mixing parameter. By using a convex combination scheme, the error signal and output signal can be calculated as follows:
\begin{subequations}
    \begin{equation}
    e(n)=\lambda(n) e_1(n) + \left[1-\lambda(n)\right]e_2(n)
    \end{equation}
    \begin{equation}
    y(n)=\lambda(n) y_1(n) + \left[1-\lambda(n)\right]y_2(n).
    \label{009}
    \end{equation}
\end{subequations}
The classical convex combination scheme adapts $\lambda(n)$ according to the \textit{sigmoid function} $\lambda(n) = \frac{1}{1+e^{-\varrho(n)}}$, where $\varrho(n)$ is the internal parameter.

At present, the convex combination strategy has been applied to both single and multi-channel ANC systems \cite{ferrer2009convex,ferrer2013convex,akhtar2020active}. To deal with the impulsive noise, the algorithm in \cite{akhtar2018normalized} introduced a convex combination to the modified FxLMP algorithm. The step size of this algorithm is based on a convex combination adaptation, which can avert the conflicted requirement between small noise residue and convergence rate.

\subsection{Fractional-order ANC algorithms}

It is worth noting that some properties of fractional-order facilitate an effective combination of the existing ANC algorithms. Some typical definitions of fractional calculus are listed in the following: Gr{\"u}nwald-Letnikov (GL), Riemann-Liouville (RL), Erd\'{e}lyi-Kober, Hadamard, Caputo, and Riesz. However, initial studies have shown that the use of GL and RL is helpful in improving noise reduction performance, particularly in the Gaussian noise source:

\textit{1) GL-based algorithms:} In \cite{aslam2015new}, a novel algorithm was proposed for the ANC system, which online estimates the secondary path according to adaptation of GL. To improve the robustness, the algorithm in \cite{shah2016fractional} exploited the output of FxLMS as the input of the fractional GL-algorithm.

\textit{2) RL-based algorithms:} The RL differintegral operator was used in \cite{shah2014fractional} to correct the updating of the parameters in the FeLMS algorithm. As shown in simulations, the new algorithm converges fast and reaches a small error.

The fractional Fourier transform (FrFT) is a unified time-frequency transform, which reflects the information in the time and frequency domains of the signal. Such transform has been applied to solve the ANC problem in \cite{durak2010adaptive}, which can obtain less error and faster convergence for various linear frequency modulated (LFM) signals.

\subsection{ANC algorithms for 3-D space}

\begin{figure}[!htb]
    \centering
    \includegraphics[scale=0.5] {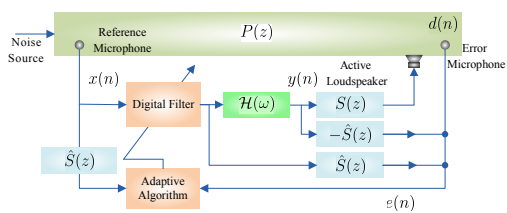}
    \caption{\label{12} Diagram of the 3-D ANC system.}
    \label{Fig12}
\end{figure}

The 3-D ANC algorithm has always been a difficult point of the technology. A novel ANC algorithm was proposed to control sound in 3-D space \cite{ardekani2014active}. The diagram of the 3-D ANC system is shown in Fig. \ref{Fig12}, where $\mathcal H(\omega)$ stands for a filter with the transfer function
\begin{equation}
\mathcal H(\omega) = \frac{3ci}{2r_{fg}\omega}j_0(\psi a \omega) + \frac{3(2r_{fg}\omega - 3ci)}{2a\psi r_{fg}\omega^2}j_1(\psi a \omega)
\label{011}
\end{equation}
where $\omega$ is the angular frequency, $r_{fg}$ is a parameter related to the locations $r_f$ and $r_g$, $a$ is the radius of the sphere, $i$ represents the imaginary unit, $c$ is the sound velocity, $\psi$ is a parameter related to the location $\phi_f$, $\phi_g$ and $c$, and $j_m$ denotes the spherical Bessel function of order $m$. On this basis, the error signal $e(n)$ can be defined in the z-domain as
\begin{equation}
e(z) \triangleq d(z) + w(z) \left(\mathcal H(z) \left(S(z) - \hat S(z)\right)+\hat S(z) \right) X(z).
\label{012}
\end{equation}
The above algorithm has demonstrated to be effectiveness in both simulations and experiments for ZoQ with arbitrary shapes (not only for spherical quiet zones). However, the 3D-ANC algorithm for multiple noise sources (such as chaotic noise and impulsive noise) is still in urgent need of research.

\subsection{Selective ANC systems}

Instead of employing the conventional real-time computation of control filter coefficients, the selective ANC (SANC) systems select the control filters from a set of pre-tuned filters based on the temporal or spectral audio features of incoming sounds. As such, the SANC systems have the improved robustness of control filters and reduced computational complexity. In \cite{ranjan2016selective}, the SANC systems were originally proposed for open window systems and then extended to other methods. In \cite{shi2019selective}, the SANC systems was integrated into the virtual microphone technique, leading to better noise reduction
performance than the conventional virtual microphone technique. More extensions of the SANC systems can be referred to \cite{shi2018novel,wen2020improved,wen2020using,shi2020feedforward}

\subsection{Distributed ANC algorithms}
\begin{figure}[!htb]
    \centering
    \includegraphics[scale=0.5] {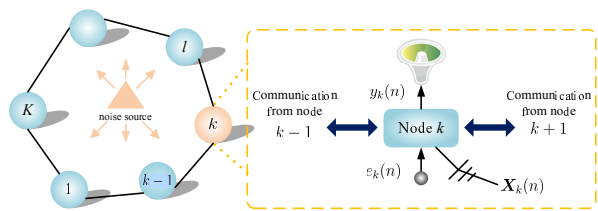}
    \caption{\label{13} Diagram of the distributed ANC system with an incremental collaborative strategy.}
    \label{Fig13}
\end{figure}
\begin{figure}[!htb]
    \centering
    \includegraphics[scale=0.5] {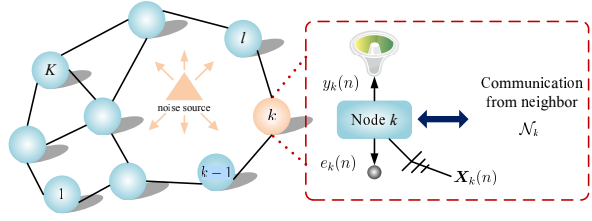}
    \caption{\label{14} Diagram of the distributed ANC system with a diffusion collaborative strategy.}
    \label{Fig14}
\end{figure}

In practice, noise source and noise cancellation points may be distanced and the nature of the noise field is complicated. Hence, it is natural to consider employing several reference microphones, error microphones and loudspeakers, resulting in multi-channel ANC system \cite{lopes2015random}. The cost of multi-channel ANC system is still expensive and demanding. The reasons are as follows. 1) To achieve both sufficient multiple coherence and time advance between the reference and error signals, a sufficient number of reference microphones are needed \cite{cheer2019application}. 2) The number of filtered reference signals is usually not equal to the number of reference signals \cite{lopes2015random}. To mitigate the computational complexity of the multi-channel ANC system, several methods were proposed, such as GPU-based methods \cite{lorente2015the,lorente2014gpu}. The theoretical behavior of the multi-channel ANC system was analyzed in \cite{ferrer2012steady-state,murata2017statistical}.  Recently, a decentralized control method by making use of frequency-domain processing was proposed for multi-channel ANC system \cite{zhang2019decentralized}.

\textit{1) Incremental algorithms:} Conventionally, the processing of multi-channel ANC systems is the \textit{centralized} estimation approach. However, it is well known that such an approach is not scalable and require restructuring of hardware components and reformulation of update rules for attenuate noise \cite{kukde2019reduced}. An effective solution is to introduce distributed ANC systems, which are motivated by distributed adaptive filtering \cite{abdolee2013estimation}. Compared with its centralized counterpart, the distributed ANC system consumes less energy and communication resources and is therefore particularly well-suited for WASNs applications. Moreover, it offers a more robust strategy that can be efficiently exploited to reduce noise over geographical regions. The earliest work using distributed ANC systems was reported in \cite{ferrer2015active}, which is based on \textit{incremental collaborative strategy}. A diagram of a distributed ANC system with an incremental collaborative strategy is shown in Fig. \ref{Fig13}, where the distributed ANC is considered with $K$ nodes, indexed by $k \in \{1,\ldots,K\}$. At time instant $n$, the observations are acquired for each noise $\{\bm X_k(n),d_k(n)\}$, where $\bm X_k(n)$ denotes the input vector, which is obtained by filtering the reference signal $\bm x(n)$ through the $k$th estimated secondary path, $d_k(n)$ denotes the noise at the sensor locations, and $e_k(n)$ is the noise residue at node $k$. The adaptation of the network for the conventional incremental FxLMS (IFxLMS) algorithm can be expressed as \cite{ferrer2015active}
\begin{equation}
\begin{array}{l}
\left\{\begin{array}{l}
\bm w_k(n) = \bm w_{k-1}(n) + \mu_k \bm X_k(n)e_k(n) \\
\bm w(n)=\bm w_K(n) \\
\end{array} \right.
\end{array}
\label{013}
\end{equation}
where $\mu_k$ is the step size at node $k$, $\bm w_k(n)$ denotes the weight vector at node $k$, and $\bm w(n)$ denotes the final estimate for iteration $n$. To further reduce computational complexity and obtain acceptable noise reduction level, a PU scheme and latency and communication constraints were introduced \cite{ferrer2015active}.

\textit{2) Diffusion algorithms:} In the incremental collaborative strategy, the definition of a cyclic path over the nodes is required, and this method is sensitive to link failure. The \textit{diffusion collaborative strategy} is a more practical strategy for WASNs applications. In this strategy, each node communicates with a subset of its neighbors $\mathcal N_k$, as shown in Fig. \ref{Fig14}. It can achieve a stable behavior over networks regardless of the topology, which is seen as a key advantage. The diffusion FxNLMS (DFxNLMS) algorithm for multi-channel ANC system was developed, whose update equation can be expressed as \cite{song2016diffusion}
\begin{equation}
\begin{array}{l}
\left\{\begin{array}{l}
\bm \varphi_k(n+1) = {\bm w}_k(n) + \mu_k \frac{\bm X_k(n)}{\left\|\bm X_k(n)\right\|^2}e_k(n) \\
\bm w_k(n+1) = \sum\limits_{l \in {\mathcal{N}_k}} a_{l,k}\bm \varphi_l(n+1) \\
\end{array} \right.
\end{array}
\label{014}
\end{equation}
where $\left\|\cdot\right\|$ is the $l_2$-norm, $\bm \varphi_l(n)$ denotes the local estimates made by neighboring nodes $l \in \mathcal N_k$, and $a_{l,k} \geq 0$ are the weighting coefficients satisfying
\begin{equation}
 a_{l,k} = 0\;\;{\rm{if}}\;\; l \notin \mathcal N_k\;\;{\rm{and}}\;\; \sum\limits_{k=1}^K {a_{l,k} = 1}.
\end{equation}

Another computationally efficient and fast adapting diffusion ANC algorithm, namely distributed FxAP (DFxAP), was proposed in \cite{ferrer2017dis}, and its convergence behavior was subsequently analyzed. To keep low complexity, the algorithm in \cite{kukde2019reduced} only considers neighbor nodes $k-1$ and $k+1$ to compute the error signal and adapt the weight vector, which provides a concise solution for practical implementation.

\section{Conclusion}
\label{sec:5}
In this paper, we have reviewed the development of linear ANC techniques in the past 10 years, with emphasis on recent methods such as sparse ANC algorithms and distributed ANC algorithms. Some fundamental frameworks of the LMS algorithm for ANC, including FxLMS, FeLMS, and FuLMS algorithms were introduced. As the non-LMS-based ANC algorithms, the type of FxAP, FxRLS, subband algorithm, and other structures ANC algorithms were investigated. It should be noted that we did not involve heuristic-based ANC algorithms in Part I, since such algorithms can also be applied to nonlinear models. Part II of this work will review NLANC techniques within the last decade, heuristic-based ANC algorithms, application of the ANC technique, and the future research challenges of ANC techniques.

\section*{Acknowledgment}
\label{sec:Ack}
The authors would like to thank the associate editor and the anonymous referees for their valuable comments.

\section*{References}

\bibliography{mybibfile}

%

\end{document}